\documentclass{aa}  
\usepackage{graphicx}
\usepackage[svgnames]{xcolor}
\usepackage{txfonts}
\usepackage{natbib}
\usepackage{hyperref}
\usepackage{soul}
\usepackage{amsmath}
\usepackage{aas_macros}
\usepackage{orcidlink}
\usepackage[normalem]{ulem}
\usepackage{multirow}

\begin{document}

   \title{Modelling the Milky Way's exoplanet population\\
   based on cosmological galaxy simulations}

   \author{Chlo\'{e} Padois\inst{1,2,3}\orcidlink{0009-0001-1380-9488}
        \and Daniel del Ser\inst{4,1}\orcidlink{0000-0001-6776-3211} 
        \and Friedrich Anders\inst{1,2,3}\orcidlink{0000-0003-4524-9363}
        \and Jo\~{a}o A. S. Amarante\inst{5,6}\orcidlink{0000-0002-7662-5475}
        \and H\'{e}lio D. Perottoni\inst{7}\orcidlink{0000-0002-0537-4146},\\
        Thomas Hajnik\inst{8}\orcidlink{0009-0001-0409-3019}
        \and Diogo Souto\inst{9}\orcidlink{0000-0001-9550-1198}
        \and Nayara I. Feliciano-Souza\inst{9}\orcidlink{0009-0001-3601-0572}
        \and Daisuke Kawata\inst{10,11}\orcidlink{0000-0001-8993-101X}
        \and Eder Martioli\inst{12, 13}\orcidlink{0000-0002-5084-168X}
}
   \institute{{Departament de F\'{i}sica Qu\`{a}ntica i Astrof\'{i}sica (FQA), Universitat de Barcelona (UB), Mart\'{i} i Franqu\`{e}s 1, 08028 Barcelona, Spain}
        \and{Institut de Ci\`{e}ncies del Cosmos (ICCUB), Universitat de Barcelona (UB), Mart\'{i} i Franquès 1, 08028 Barcelona, Spain}
        \and{Institut d'Estudis Espacials de Catalunya (IEEC), Edifici RDIT, Campus UPC, 08860 Castelldefels (Barcelona), Spain}
        \and{Observatori Fabra, Reial Acadèmia de Ci\`{e}ncies i Arts de Barcelona, Rambla dels Estudis, 115, E-08002 Barcelona, Spain}
        \and{Department of Astronomy, School of Physics and Astronomy, Shanghai Jiao Tong University, 800 Dongchuan Road, Shanghai, 200240, China}
        \and{State Key Laboratory of Dark Matter Physics, School of Physics and Astronomy, Shanghai Jiao Tong University, Shanghai, 200240, China}
        \and{Observat\'{o}rio Nacional, MCTI, Rua Gal. Jos\'{e} Cristino 77, Rio de Janeiro, 20921-400, RJ, Brasil}
        \and{Institute of Astronomy, University of Cambridge, Madingley Road, Cambridge, CB3 0HA, United Kingdom}
        \and{Universidade Federal de Sergipe, Av. Marechal Rondon, S/N, 49000-000 São Crist\'{o}v\~{a}o, SE, Brazil}
        \and{Mullard Space Science Laboratory, University College London, Holmbury St. Mary, Dorking, Surrey RH5 6NT, UK}
        \and{National Astronomical Observatory of Japan, 2-21-1 Osawa, Mitaka, Tokyo 181-8588, Japan}
        \and{Laborat\'{o}rio Nacional de Astrof\'{i}sica, Rua Estados Unidos 154, 37504-364, Itajub\'{a} - MG, Brasil}
        \and{Instituto Nacional de Pesquisas Espaciais/MCTI, Av. dos Astronautas, 1758, S\~{a}o Jos\'{e} dos Campos, SP, Brazil}
    }

   \date{Received 2 September 2025; accepted 13 November 2025}

    \abstract
    {Exoplanet transit and radial-velocity surveys have allowed us to explore the exoplanet population in our Galactic surroundings. The planet populations in more remote areas of the Milky Way (MW) will become accessible with future instrumentation.}
    {In this paper we aim to simulate realistic exoplanet populations across different regions of the MW by combining state-of-the-art cosmological simulations of our Galaxy with exoplanet formation models and observations.}
    {We model the exoplanet populations around simulated single stars, using planet occurrence rate and multiplicity depending on stellar mass, metallicity, and planet type, and assign them physical parameters such as mass and orbital period.}
    {Focussing first on the solar vicinity, we find mostly metallicity-driven differences in the distributions of non-hosting and planet-hosting single stars. 
    In our simulated solar neighbourhood, 52.5\% of all planets are Earth-like (23\% of them located in the Habitable Zone), 44\% are super-Earths/Neptunes, and $3.5\%$ are giant planets.
    A detailed comparison with the census of \textit{Kepler} exoplanets and candidates shows that, when taking into account the most relevant selection effects, we obtain a similar distribution of exoplanets compared to the observed population.
    However, we also detect some significant differences in the exoplanet and host star distributions (e.g. more planets around F-type and red-giant stars compared to the observations) that we attribute mostly to a too strong recent star formation and a too large disc scale height in the simulation compared to the solar neighbourhood, as well as to some caveats in our exoplanet population synthesis that will be addressed in future work. 
    Extending our analysis to other regions of the simulated MW and to other galaxies within the same suite of simulations, we find that the relative percentages of Earth-like, super-Earth/Neptunians, and giant planets remain largely consistent as long as the simulated galaxy matches the morphology and mass of the MW.}
    {We have created a fast and flexible framework to produce exoplanet populations based on MW-like simulations that can easily be adapted to produce predictions for the yields of future exoplanet detection missions.}
  

\maketitle

\section{Introduction \label{sec:intro}}

The discovery, exploration, and study of exoplanets have started a new era in astrophysics. After 30 years of discoveries, the field of exoplanetary science has reached remarkable advancements in recent years through innovative observation techniques, including better precision in radial velocity measurements, transmission spectroscopy, or direct imaging. As of February 2025, 5823 exoplanets have been confirmed\footnote{\url{https://exoplanetarchive.ipac.caltech.edu/}} \citep{NASAexoplanetaryarxiv}. 

Several space missions have already explored the exoplanet population in our Galactic surroundings. The most successful surveys so far have been the {\it Kepler} Space Telescope (\citealt{Borucki2010}) and its follow-up mission K2 \citep{Howell2014}, which already account for more than half of the currently confirmed exoplanets (\citealt{Lissauer2024}, \citealt{Mayo2018}), the Transiting Exoplanet Survey Satellite (TESS; \citealt{Pal2018}), and the CoRoT mission \citealt{Baglin2006, Deleuil2018}. Presently, the progressive processing of the {\it Gaia} mission data (\citealt{GaiaCollaboration2016, GaiaCollaboration2023Vallenari}) is yielding more and more astrometric detections of exoplanets (e.g. \citealt{Holl2023, Sahlmann2025}), ESA's CHaracterising ExOPlanet Satellite (CHEOPS, \citealt{Rando2018, Fortier2024}) is characterising known exoplanets and their host stars, while the NASA flagship mission James Webb space telescope (JWST, \citealt{McElwain2023}) is already carrying out transmission spectroscopy of possibly habitable Earth-like exoplanets (e.g. \citealt{Ahrer2023}). The Planetary Transits and Oscillations of stars mission (PLATO; \citealt{Rauer2014, Rauer2025}) and Ariel (Atmospheric Remote-sensing Infrared Exoplanet Large-survey;\citealt{Tinetti2022}) are next-generation telescopes that will provide crucial new statistics on the frequency of different types of exoplanetary systems around different types of stars. 

None of the exoplanetary systems discovered to date resemble our solar system, with almost 80\% of the confirmed exoplanet\footnote{\url{https://exoplanetarchive.ipac.caltech.edu/docs/counts_detail.html}} with measured mass being heavier than $10\rm ~M_{\oplus}$, while half of the Solar System is composed of small planets.
This may be explained by the limitations of the current instruments (in particular, transits and radial-velocity surveys), which tend to favour the detection of massive, short-period planets. Current surveys that account for these selection biases indicate that small planets are more common than giant planets \citep{Johnson2010, Dressing2013, Burke2015, Bryson2021}. To compensate for the observational biases, various studies tried to derive the probability of a star to host exoplanets (i.e. the occurrence rate), depending on the exoplanet type and the star's properties, mainly the stellar mass and metallicity (see \citealt{Pan2025} and references therein).
For example, the occurrence rate of giant planets depends on various variables: it tends to increase with longer orbital periods, as well as with the host star's metallicity and mass \citep[e.g.][]{Fischer2005, Petigura2018}.
It has also been observed that multiplanetary systems with resonant orbits and size similarities are common and that exoplanets can form and survive in multiple stellar systems (e.g. \citealt{Cochran1997, Anglada-Escude2016, Weiss2018, Triaud2022}).

This work aims to simulate a simple but realistic exoplanet population by modelling the main known factors that influence the existence and distribution of different types of exoplanets. We start from a cosmological simulation of a Milky Way-like galaxy, transform star-particles into stars using stellar evolutionary models, and then generate a synthetic exoplanet population by joining observational results with predictions from exoplanet formation models. Our predictions can thus provide a benchmark for future exoplanet detection missions such as PLATO, HAYDN (High-precision AsteroseismologY in DeNse stellar fields; \citealt{Miglio2024}), or the {\it Roman} Space Telescope's Galactic Bulge Time Domain Survey \citep{Wilson2023}.

Our paper is structured as follows: In Sect. \ref{sec:method} we describe our method of creating stellar and exoplanetary populations from cosmological galaxy simulations. In Sect. \ref{sec:results_SN} we present the results for a volume-complete exoplanet census of the solar vicinity in our fiducial simulation. In Sect. \ref{sec:kepler} we compare our model with observations from {\it Kepler} using a detailed forward simulation that includes the most relevant selection effects. Sect. \ref{sec:same_glx_different_regions} then extends our modelling to different environments in the same galaxy, while Sect. \ref{sec:different_glx_same_region} studies the differences among the simulated solar-neighbourhood exoplanet populations in six simulated galaxies. In Sect. \ref{sec:discussion} we summarise and discuss our results in light of the existing literature, and discuss possible future applications and improvements.

\section{Methodology}\label{sec:method}

In this work, we aim to simulate a realistic population of exoplanets, starting from a Milky Way-like cosmological galaxy simulation. The star particles used to simulate the initial population of host stars are extracted from a well-tested and publicly available snapshot of a high-resolution cosmological simulation (Sect. \ref{sec:simulation}). Focussing on a similar environment to our Solar System in terms of location in the Milky Way, we generate stellar systems from the stellar particles of the host simulation, using a set of stellar models, initial mass function, and multiplicity statistics (Sect. \ref{sec:stars}). Finally, we create planets around these synthetic stars using a series of prescriptions inspired by both the observed exoplanet census and planet population synthesis models (Sect. \ref{sec:exosim}). 

\subsection{Cosmological simulation and solar neighbourhood}\label{sec:simulation}

\begin{figure}
    \centering
    \includegraphics[width=0.49\textwidth]{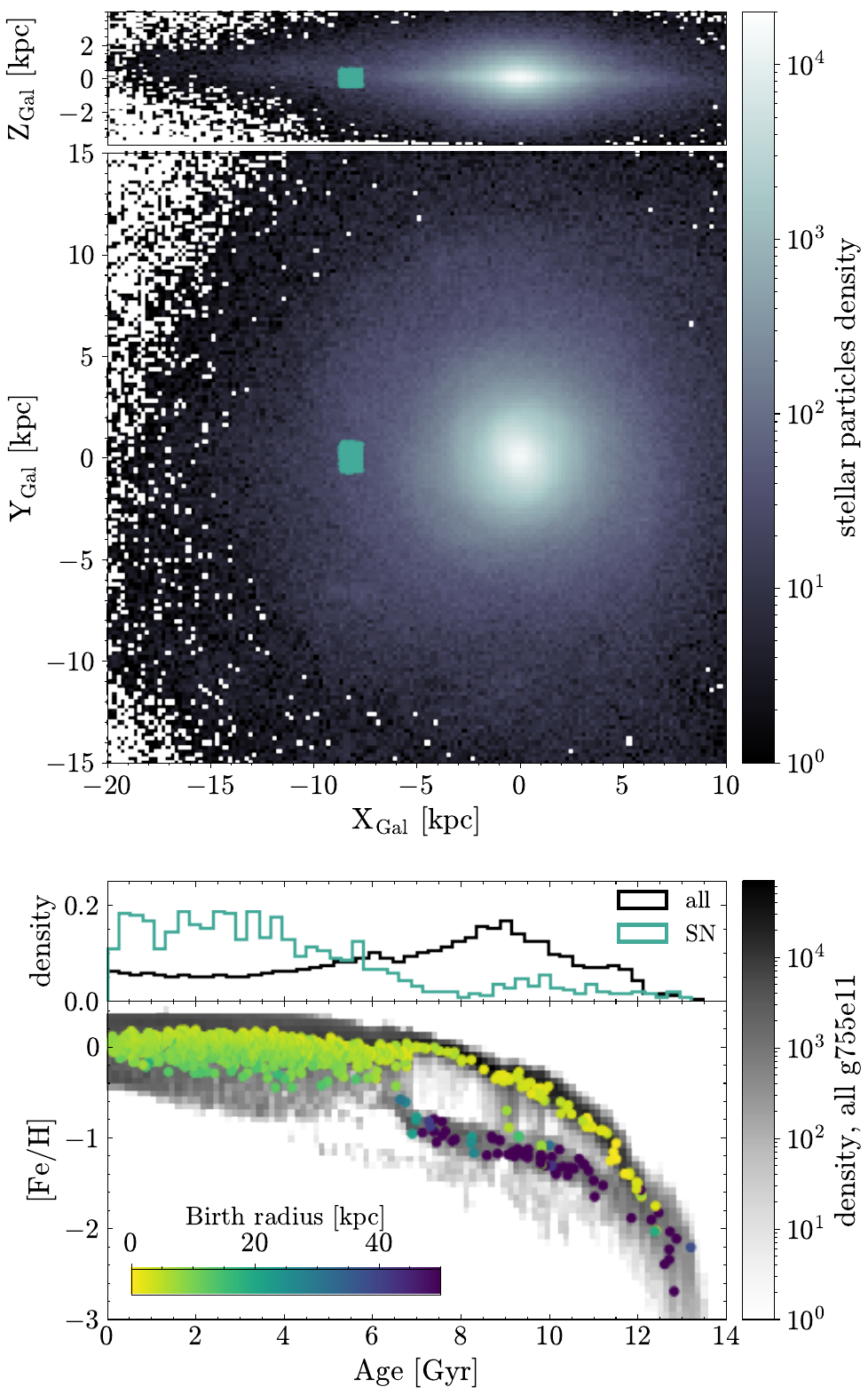}
    \caption{Selection of the stellar particles of the NIHAO-UHD simulation g7.55e11 snapshot 1024 (\citealt{Buck2020}). Top panel: Stellar particles density (number count per pixel) in Galactocentric Cartesian coordinates (in edge-on and top-down views), highlighting the selected solar neighbourhood (1147 stellar particles, highlighted in turquoise). Bottom panel: Age distribution and age-metallicity relation of the selected particles, coloured by Galactocentric birth radius. Particles not belonging to the SN selection are shown in grey as reference.}
    \label{fig:SN_selection_and_FEH_with_age}
\end{figure}

We use Numerical Investigation of a Hundred Astronomical Objects - Ultra High Definition (NIHAO-UHD) suite of cosmological hydrodynamical simulations \citep{Buck2020}, which comprises six cosmological zoom-in simulations of Milky Way-mass galaxies from the NIHAO simulation suite \citep{Wang2015} resimulated at higher resolution (mean stellar mass of $\sim 9300~ M_{\odot}$). For the purpose of this work, we use the stellar particles of the publicly available final snapshots of each of these galaxies. We begin by focussing on one of them, {\tt g7.55e11}, and show its face-on and edge-on views in the top panel of Fig.~\ref{fig:SN_selection_and_FEH_with_age}. {\tt g7.55e11} displays several features similar to the MW: total stellar mass, stellar disc size, rotation curve, double exponential vertical disc profile (see \citealt{Buck2020}), and also has a bimodality in the [$\alpha$/Fe] vs. [Fe/H] in the solar neighbourhood (see \citealt{Buck2020b}), similar to what is observed in the MW (e.g. \citealt{Fuhrmann1998, Anders2014, Hayden2015, Queiroz2020, Imig2023}). It also offers a higher resolution in stellar particles, compared to \texttt{g8.26e11} for example, which would also be a good candidate from the NIHAO-UHD suite of simulations. It does not, however, match the star-formation history of the solar vicinity very well (too many young stars are found in {\tt g7.55e11} relative to the solar neighbourhood census (e.g. \citealt{Mazzi2024, Gallart2024, delAlcazar-Julia2025}). This difference is especially visible when studying the \textit{Kepler} field of view in Sect. \ref{sec:kepler}, when selecting the observable stars. In Sect. \ref{sec:different_glx_same_region} we provide a comparison with the other galaxies of the NIHAO-UHD simulation suite.

In this paper, we focus on the (extended) solar neighbourhood (SN), where most of the presently known exoplanets reside. Since this simulated galaxy has a similar size to the Milky Way, we selected stellar particles with Galactocentric cylindrical coordinates comprised between 7.7 kpc $< {\rm R}_{\rm Gal} < 8.7$ kpc, $|{\rm Z}_{\rm Gal}| < 0.5$ kpc, $\phi = (30\pm5)^{\circ}$ (i.e. a region lagging behind the Galactic bar, based on the value favoured by \citealt{Bland-Hawthorn2016}). This selection is illustrated in the upper panel of Fig. \ref{fig:SN_selection_and_FEH_with_age} and results in 1147 stellar particles that can be interpreted as simple stellar populations of $5-9 \cdot 10^3~M_{\odot}$ each (a typical value for star-forming complexes in the Milky Way; e.g. \citealt{Negueruela2025}).
The bottom panel of Fig. \ref{fig:SN_selection_and_FEH_with_age} shows the age-metallicity relation of the selected solar-vicinity volume as coloured points. The bimodality for ages $\gtrsim 7$ Gyr (corresponding to birth radii $>20$ kpc) arises from the accretion of a smaller satellite galaxy, similar to the {\it Gaia} Enceladus/Sausage merger that the Milky Way experienced $\gtrsim9$ Gyr ago \citep{Belokurov2018, Helmi2018, Gallart2019, Limberg2022}.

\subsection{From star particles to stellar systems}\label{sec:stars}

To translate star particles into individual stars, we create simple stellar populations from a set of stellar evolutionary models. In this work, we use the PARSEC 1.2S + COLIBRI models\footnote{\url{https://stev.oapd.inaf.it/cgi-bin/cmd}} \citep{Bressan2012, Chen2015, Marigo2017, Pastorelli2020} with standard choices regarding mass loss ($\eta_{\rm Reimers}=0.2$) and YBC bolometric corrections \citep{Chen2019} with the revised Vega spectrum from \citet{Bohlin2020}. These models use scaled-solar composition with a $Z$-dependent helium content, microscopic diffusion, nuclear reaction rates from the JINA REACLIB database \citep{Cyburt2010}, the FREEEOS\footnote{\url{https://freeeos.sourceforge.net/}} equation of state, OPAL opacities \citep{Iglesias1996} except in the low-temperature regime, and a mixing-length parameter \citep{Bohm-Vitense1958} of $\alpha_{MLT}=1.74$.

In practice, for each particle we select the closest metallicity and age in a PARSEC grid of evolutionary models, and then sample a total number of stars (at any evolutionary stage) corresponding to the particle's mass in the cosmological simulation from an initial mass function. These synthetic single stars are then arranged in stellar systems according to the observed multiplicity statistics  (see Sect. \ref{annexe:generate_multiple_stellar_syst}).

The stars simulated in this work were generated following the initial mass function (IMF) determined by \citet{Kirkpatrick2024} using a 10 pc volume-complete sample of stars in the solar vicinity.
After creating all the stars in the stellar particles we make a cut in mass, from 0.1 to 1.6 $\rm M_{\odot}$ in order to keep only stars of type F, G, K and M. O, B, and A stars were excluded because of their short lifetimes and the difficulty in detecting and characterising exoplanets in these systems.

Binary stars are often overlooked as planetary hosts in similar studies (e.g. \citealt{Madau2023, Boettner2024}), as they are systems with complex processes that exhibit a wide variety of instabilities, frequently impacting the stability of the planetary system. Nearby examples, such as the Proxima Centauri system, as well as numerous recent discoveries via transit-timing variations \citep[e.g.][]{Doyle2011, Kostov2013, Anglada-Escude2016, Getley2017, Kostov2020}, radial-velocity searches (e.g. \citealt{Martin2019}), microlensing \citep{Bennett2016}, and imaging companion searches around known planet hosts \citep{Sigurdsson2003, Mugrauer2025}, show that planets often reside in multiple systems (for recent reviews see \citealt{Kostov2023}, \citealt{Deeg2024}), orbiting the whole multiple system (``P-type'' or circumbinary planets, for binary systems) or one of its components (``S-type'' planets).
However, for the purpose of this paper, we focus exclusively on single stars, not considering exoplanetary systems involving binaries, triples or higher-order multiple stellar systems, nor stellar remnants (white dwarfs, neutron stars, and black holes), due to poor constraints in the present-day exoplanet census.
We still include binary and higher-order systems in our stellar population synthesis to obtain a realistic stellar population (see Appendix \ref{annexe:generate_multiple_stellar_syst}.)
This assumption may lead to an underestimation of the number of exoplanets, as recent studies estimate that around 20\% of detected planets orbit multiple systems (see \citealt{Thebault2025} for an extensive analysis of the planet-hosting multiple systems). We will include S-type exoplanets in future work.

\subsection{Simulating exoplanets \label{sec:exosim}}

To simulate the exoplanet population orbiting around the simulated stellar systems, we use occurrence rates of different planetary types derived from the literature. We assign to each single star, depending on its mass and metallicity, a number of Earth-like planets, super-Earth (SE)/Neptunian planets, and sub-giants/giants (see Sect. \ref{sec:occurrence}). Then we assign a mass and a period to each planet, drawn from bivariate Gaussian distributions obtained by combining observations and planetary formation models (see Sect. \ref{sec:mass-period}). The orbital eccentricity is considered to be zero for all planets, as a first approximation. 

\subsubsection{Occurrence rates and multiplicity}\label{sec:occurrence}

\begin{figure}
    \centering
    \includegraphics[width=0.49\textwidth]{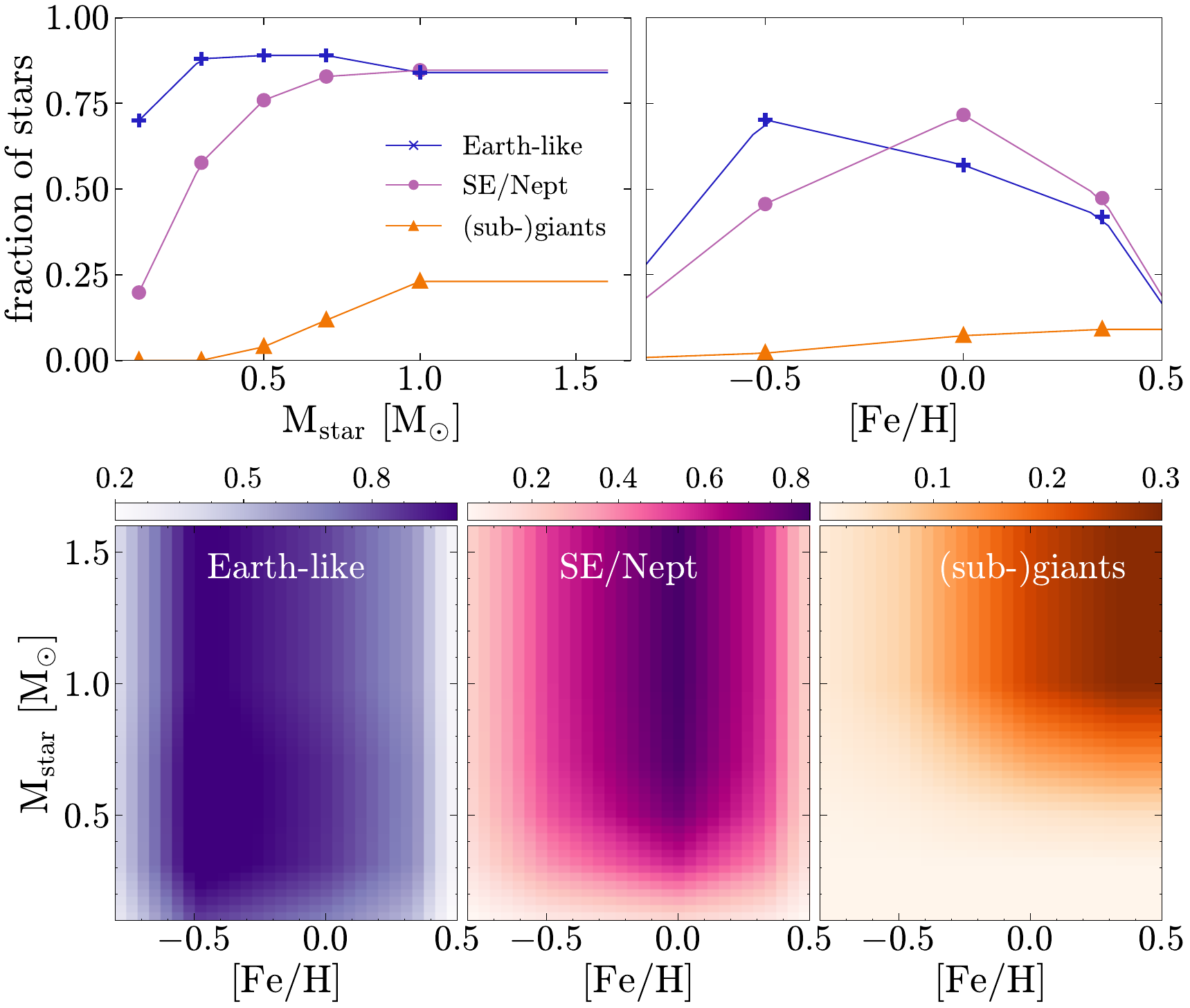}
    \caption{Fractions of single stars harbouring at least one planet of different planetary types. The top left panel shows the occurrence rate as a function of the stellar mass, adapted from \cite{Burn2021}. 
    The top right panel shows the occurrence rates as a function of the stellar metallicity, adapted from \cite{Narang2018}. The bottom row shows the joint occurrence rates obtained for each category of planet, combining the stellar mass and metallicity dependencies.}
    \label{fig:occ_rate_planets_mass_feh}
\end{figure}

Similarly to stellar multiplicity, the occurrence rates and multiplicities of different planetary types depend on the stellar properties (see, e.g. Sect. 2 of \citealt{Drazkowska2023} for a review). In this work, we consider only the dependence on stellar mass and metallicity because these are the most established dependencies. 
However, there are also indications for additional dependence of planet statistics and composition on stellar age (e.g. \citealt{Weeks2025}) and detailed chemical composition (e.g. \citealt{Adibekyan2012, Tautvaisiene2022, Cabral2023, Sharma2024, Yun2024}).
The occurrence rates as a function of stellar mass and metallicity used in this paper (see below) for the three categories (Earth-likes, super-Earths/Neptunians, sub-giants/giants) are presented in Fig. \ref{fig:occ_rate_planets_mass_feh}, as well as the global occurrence rate obtained by combining those two dependencies.

For the mass dependence, we adapt results from \cite{Burn2021}: from a planet formation model (based on the Generation III Bern model; detailed in \citealt{Emsenhuber2021a}) they derive the fraction of stars harbouring at least one planet of five different types (Earth-like, super-Earth, Neptunians, sub-giants, and giants; see their section 3.2), as a function of the stellar mass. We group together the super-Earth and Neptunian planets, as well as the sub-giants and giants, to end up with three categories. As \cite{Burn2021}'s simulations are only available for stars with masses up to $1~\mathrm{M}_{\odot}$ we use constant extrapolation to extend to our maximum mass of $1.6~\mathrm{M}_{\odot}$.

There is ample evidence from observations as well as simulations for a dependence of planet occurrence rates on metallicity (e.g. \citealt{Gonzalez1997, Fischer2005, Mordasini2012, Narang2018}). In general terms, there is a consensus that the occurrence rate of giant planets strongly depends on metallicity, while the occurrence rate of rocky planets does not significantly vary with metallicity above a certain metallicity floor. There is, however, no consensus on the value of this metallicity floor (e.g. \citealt{Andama2024}).

In this work, we use the results derived by \citet{Narang2018} from \textit{Kepler} DR25 results to describe the metallicity dependence of the planet occurrence rate (Fig. \ref{fig:occ_rate_planets_mass_feh}, upper right panel). To match our three planet categories, we combine their data for the $2-4 \rm ~R_{\oplus}$ and $4-8 \rm ~R_{\oplus}$ planets in one group, corresponding to our super-Earths/Neptunians category. We conservatively set the occurrence rate to zero for metallicity lower than $-1.0$ and higher than 0.6, reproducing the lower and upper limits of the detected host-star metallicity distribution. 

\begin{figure}
    \centering
    \includegraphics[width=0.49\textwidth]{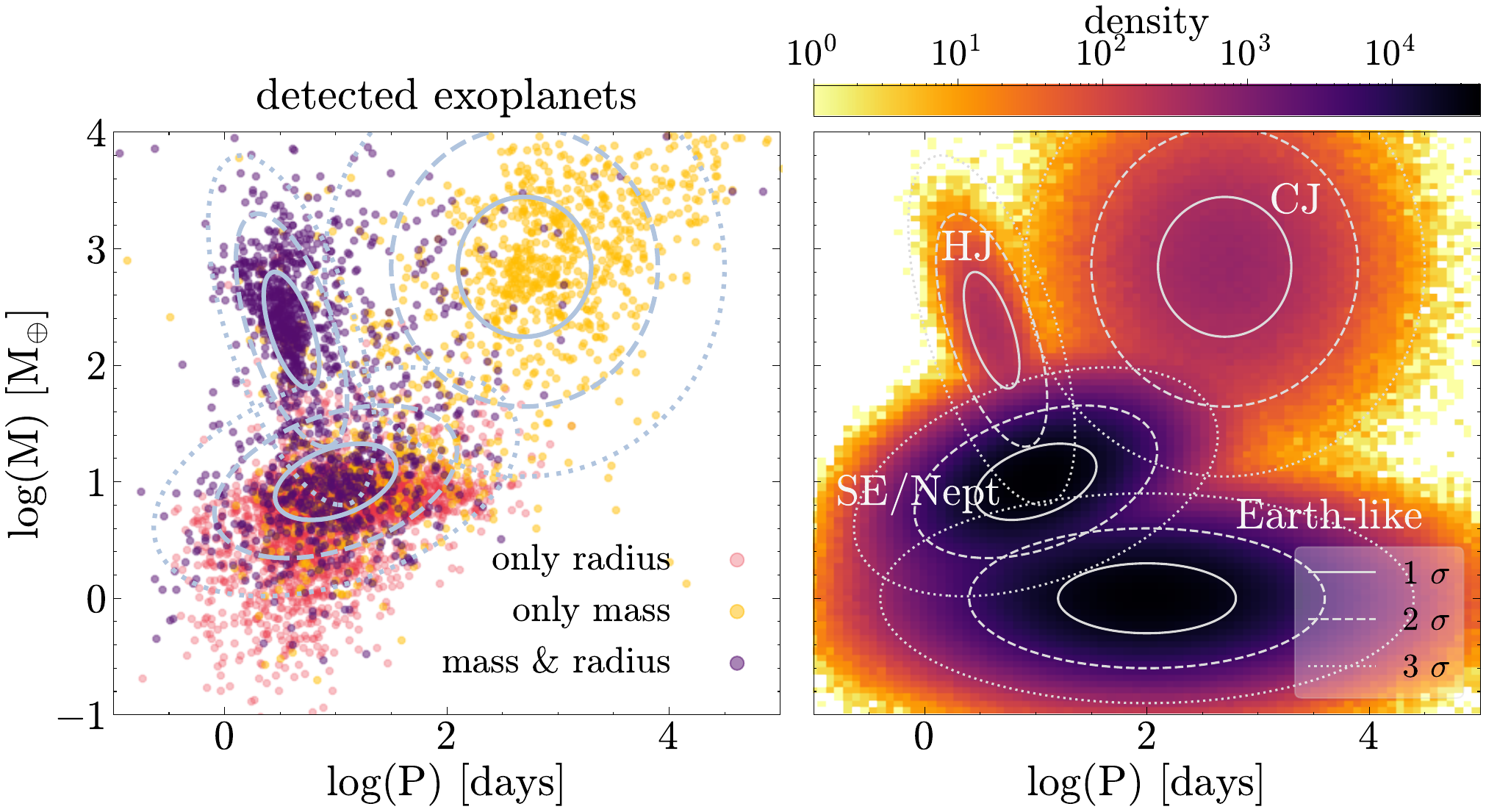}
    \caption{Distribution of detected vs. simulated exoplanets. {\it Left panel:} Detected exoplanets from the NASA Exoplanet Archive. Planets with both mass and radius measurements are shown in purple, and the ones with only mass measurements are in yellow. Planets with only radius measurement are in red (using the radius-to-mass conversions of \citealt{Parc2024}), but were not used to derive the bivariate Gaussian distributions. {\it Right panel:} Density distribution (number count per pixel) of the 24.5 million simulated exoplanets in the solar vicinity. In grey lines are overplotted the bivariate Gaussian distributions used to assign a mass and a period to each planet randomly. The distribution of giants (HJ and CJ) and SE/Neptunians were chosen to reproduce the confirmed exoplanet population (see left panel) while the Earth-like distribution was based on the distribution predicted by \cite[][their Fig. 13]{Drazkowska2023}. }
    \label{fig:MP_fit_Drazkowska}
\end{figure}

\begin{table}
\caption{Planet types definitions, based on \cite{Burn2021} for masses and on \cite{Narang2018} for radii.}
\label{tab:planet_types_boundaries}
\centering
\begin{tabular}{c c c}
\hline\hline
planet type & mass boundaries & radius boundaries \\
\hline
    Earth-likes & $M < 2~M_{\oplus}$ & $R < 1.25~R_{\oplus}$ \\
    SE/Neptunes & $2 \le M < 30~M_{\oplus}$ & $1.25 \le R < 4~R_{\oplus}$ \\
    sub-giants/giants & $M \ge 30~M_{\oplus}$ & $R \ge 4~R_{\oplus}$ \\
\hline
\end{tabular}
\end{table}

To combine the dependencies of the planet occurrence rate on the stellar mass and metallicity, we assume that they are independent. 
We multiply the occurrence rate as a function of stellar mass with the occurrence rate as a function of [Fe/H], normalised such that we match the occurrence rate of \citet{Burn2021} for stars with [Fe/H]$~=0$
(see Fig. \ref{fig:occ_rate_planets_mass_feh}). When we obtain a combined occurrence rate greater than 1, we consider it equal to 1. 

For our single stars sample, we draw their probability of harbouring at least one planet of each planet type using the combined occurrence rate. Then we draw the number of planets of each type using the multiplicity as a function of the stellar mass given by \cite{Burn2021}, combining again their 5 categories in 3 by simple addition. The number of planets of each category is drawn randomly following a Poisson distribution centred on the multiplicity value at their stellar mass.

\subsubsection{Mass-period distributions}\label{sec:mass-period}

We then assign fundamental astrophysical parameters to all created planets, in particular mass and orbital period.
The distribution of confirmed exoplanets from the NASA Exoplanet Archive \citep{Akeson2013, NASAexoplanetaryarxiv} shows clear observational biases, with an over-representation of hot Jupiters and a lack of Earth-like planets, due to the large detectability dependence on planetary radius and orbital period (Sect. \ref{sec:detectability}). We aim to account for this bias by combining the observed distribution with predictions from a planetary formation model. 
We approximate the Mass-Period distribution of the different planet categories by bivariate Gaussians.
For the super-Earth/Neptunian planets and the giants, the bivariate Gaussians parameters were chosen to reproduce the observational data (confirmed exoplanets from the NASA Exoplanet Archive). 
For the Earth-like population, we adjusted the bivariate Gaussian parameters such that they reproduce the combination of the distributions shown in
\citet[][both panels of their Fig. 13]{Drazkowska2023}.\footnote{
These authors obtain the first distribution with their own simplified pebble accretion population synthesis (their Sect. 5.2), assuming that all pebbles are converted into planetesimals (their Sect. 5.4) and using 300 m and 50 km planetesimals. The second distribution is the result of the Bern single-planet synthesis model \citep{Emsenhuber2021b}, using 300 m planetesimals.}
The four bivariate Gaussian distributions are overplotted as light grey contours in Fig. \ref{fig:MP_fit_Drazkowska}, and their parameters can be found in Table \ref{tab:bivariate_gaussian_params}

We assign a random mass and orbital period to all planets, sampling from the bivariate Gaussian associated with their assigned type. For the sub-giants/giants category, we randomly distribute them between hot Jupiters (``HJ'') and cold/warm Jupiters (``CJ''), with a 10\% probability of being a HJ.
We redraw the parameters for planets with masses greater than $10^4~\rm M_{\oplus}$ (the lower mass limit of brown dwarfs; e.g. \citealt{Burrows1993, Joergens2014}) as well as planets with a distance to their host star lower than two times the stellar radius. 

\begin{figure}[h]
    \centering
    \includegraphics[width=0.49\textwidth]{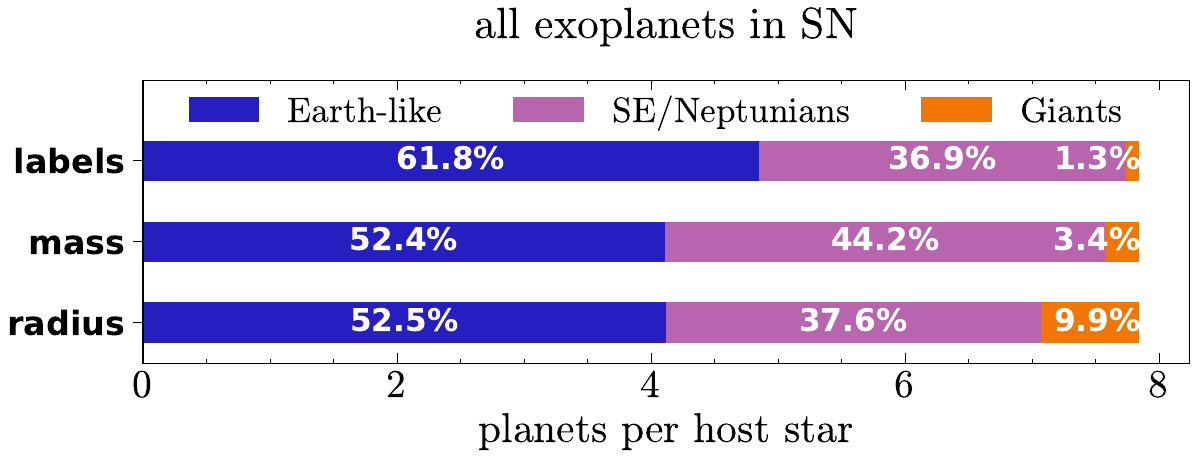}
    \caption{All simulated exoplanets in the volume-complete SN. Influence of the planet type definition on the obtained occurrence rate, comparing three different definitions for the exoplanet categories: using the original labels, the mass or the radius.}
    \label{fig:SN_percents_pl_types_variousdefs}
\end{figure}

In what follows, we will use these assigned planet masses to classify the exoplanets into the different planetary types (see the boundaries used in Table \ref{tab:planet_types_boundaries}), instead of the original parent Gaussian in Fig. \ref{fig:MP_fit_Drazkowska}, which could be incompatible with the drawn mass value. A priori, we can think of several different ways to classify exoplanets into planet types, for example by mass, by radius, or by density. In Fig.~\ref{fig:SN_percents_pl_types_variousdefs} we compare the percentages obtained for the volume-complete SN (see Sect. \ref{sec:results_SN}) using three different classifications: 1. \texttt{``labels''}, using the planet type first assigned using the combined occurrence rate of the different planet types (see Sect.~\ref{sec:occurrence} and Fig.~\ref{fig:MP_fit_Drazkowska}); 2. \texttt{``mass''}; and 3. \texttt{``radius''} using the boundaries set in Table~\ref{tab:planet_types_boundaries}.
As we assign masses drawing from broad bivariate Gaussian distributions, the associated mass can end up being incoherent with the initial label, leading to differences in the percentages of each planet type. 
To achieve consistent classifications, we use mass-based categories for the simulated exoplanets and radius-based definitions for the \textit{Kepler} exoplanets (since \textit{Kepler} exoplanets are detected by transit, they have accurate radius measurements instead of mass). It would be ideal to use the same category definition for simulated and observed planets, but there are not enough accurate mass measurements for planets detected by \textit{Kepler}. Thus, our conversion from masses to radii for simulated planets adds another important uncertainty.

In addition, we do not correct for potential overlaps of orbits within a planetary system, nor do we consider the systems' dynamic stability.
In Fig. \ref{fig:planetary_systs} we show
the architecture (planet distribution vs. semi-major axis) of 40 randomly selected synthetic planetary systems, sorted by stellar mass. We insist that we do not currently aim to produce realistic planetary architectures (see, e.g. \citealt{Howe2025}), but only try to reproduce the overall statistics of the observed exoplanet population (see Sect. \ref{sec:kepler} for a comparison with the {\it Kepler} field). Dedicated analyses of the variety of exoplanet architectures found by {\it Kepler} (e.g. \citealt{Fabrycky2014, Lissauer2024}), CHEOPS, and TESS indeed reveal signs of orbital resonances \citep{Luque2023, Dai2024}, instabilities \citep{Ghosh2024}, dynamical packing \citep{Obertas2023}, etc. that are out of the scope of this work. 
The exoplanetary population obtained in this work is only expected to show realistic properties in terms of metallicity, effective temperatures, radius of the planets, and semi-major axis of their orbits.
\cite{Mishra2021} suggested to divide the planetary systems' architecture into 4 classes: similar, anti-ordered, ordered, and mixed. Without imposing any constraint on the system's architecture (not correcting orbits overlaps, not considering resonances), we obtain similar frequencies to the Bern model used by \cite{Mishra2021}: see lower panel of Fig. \ref{fig:planetary_systs}.

\section{Results for the volume-complete solar neighbourhood}\label{sec:results_SN}

In this section we describe our results obtained for a volume-complete solar neighbourhood (of approximatively $1 ~\rm pc^3$) in our fiducial galactic simulation. We compare the simulated exoplanets to observed exoplanets in Sect. \ref{sec:kepler}.

\subsection{Simulated stellar population}
\begin{figure}
    \centering
    \includegraphics[width=0.49\textwidth]{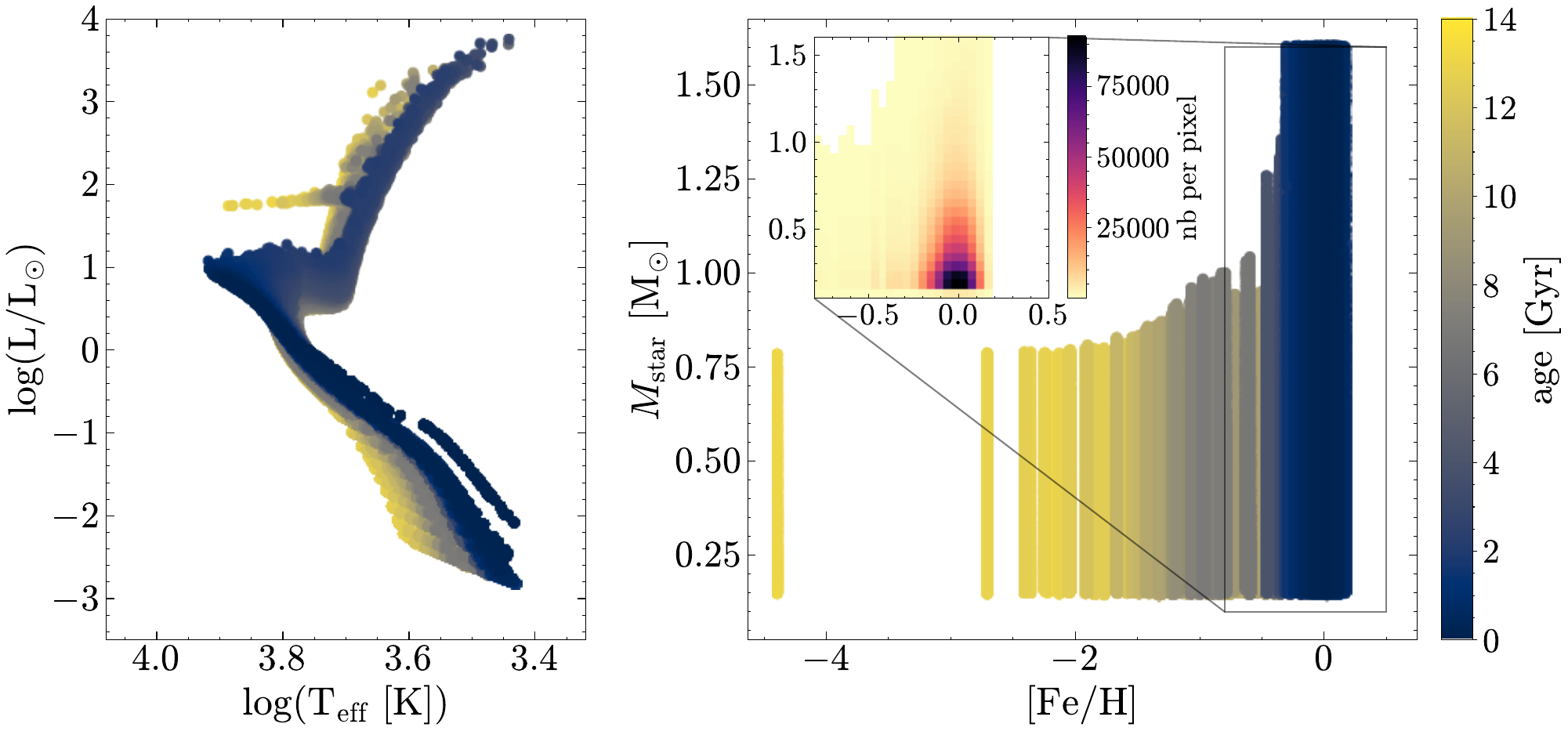}
    \caption{All simulated stars in the volume-complete SN of the \texttt{g7.55e11} galactic simulation. Left panel shows the Hertzsprung–Russell diagram, colour coded by age. Right panel shows the distribution of stellar mass versus metallicity, also colour coded by age. The zoom-in panel shows the density distribution of stars in the region of the mass-metallicity space plotted in Fig. \ref{fig:occ_rate_planets_mass_feh}.}
    \label{fig:HR_massFeH_SN_coloredage}
\end{figure}

As described in Sect. \ref{sec:simulation}, we start by selecting the 1147 stellar particles corresponding to the SN in the galaxy \texttt{g7.55e11}'s redshift-zero snapshot (Fig. \ref{fig:SN_selection_and_FEH_with_age}). From the stellar particles, we create 9.1 million individual stars, from which we keep only FGKM stars (Sect. \ref{sec:method}), excluding stars with mass greater than 1.6 $\rm M_{\odot}$: we obtain a total sample of around 8.3 million stars. After the creation of stellar systems as detailed in Appendix \ref{annexe:generate_multiple_stellar_syst}, we obtain around 3.4 million single stars, 3.6 million in binary systems, and 1.3 million in higher-order systems. The total stellar density in our SN selection is around $0.005 \rm ~M_{\odot}\cdot pc^{-3}$, while the latest estimation based on observations gives results around $0.04 \rm ~M_{\odot}\cdot pc^{-3}$ (\citealt{Lutsenko2025}). The Hertzsprung–Russell diagram and the distribution in mass-metallicity diagram are shown in Fig. \ref{fig:HR_massFeH_SN_coloredage}.

\subsection{Exoplanet demographics in the simulated solar vicinity}

Simulating exoplanets only around single stars as described in Sect. \ref{sec:exosim}, we end up with a total sample of 22.6 million planets, organised in 2.9 million planetary systems and composed of around 52.5\% of Earth-like planets, 44\% of super-Earths and Neptunians, and around 3.5\% of giants and sub-giants. About 23\% of them are located in the circumstellar habitable zone (CHZ; see Sect. \ref{sec:habitability}). The distribution of the exoplanet sample in the mass-period diagram is shown in Fig. \ref{fig:MP_fit_Drazkowska}. We stress that, by construction (see Table \ref{tab:bivariate_gaussian_params}), our mass-period diagram reproduces the observed data for cold and hot Jupiters, as well as super-Earths and Neptunes (as reported in Fig. 13 of \citealt{Drazkowska2023}).

\begin{figure*}
    \centering
    \includegraphics[width=0.99\textwidth]{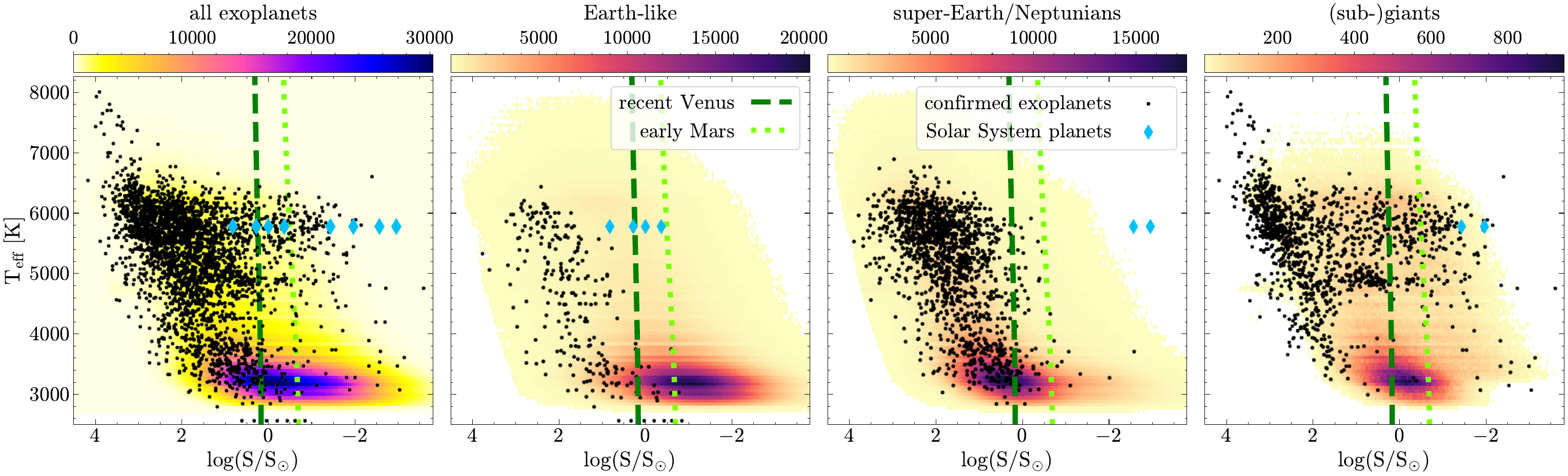}
    \caption{Effective temperature of the host stars vs. instellation at the simulated exoplanet for the volume-complete SN sample. For comparison, we overplot the distribution of the observed exoplanets \citep{NASAexoplanetaryarxiv} as black dots and the solar-system planets as cyan diamonds. The left panel shows the entire sample, while the other three panels show the distribution of each planet category (Earth-like, SE/Neptunians, giant planets). The limits we adopted for the definition of the CHZ are plotted in dark and light green (see Sect. \ref{sec:habitability}).}
    \label{fig:teff_instellation}
\end{figure*}

Fig. \ref{fig:teff_instellation} shows the effective temperature vs. instellation (the intensity of the received star's radiation, in surface power density) diagrams for our simulated solar vicinity sample, compared to all currently confirmed exoplanets in the NASA exoplanet database.
The left panel of Fig. \ref{fig:teff_instellation} contains the full sample of planets, while the other panels correspond to our sub-categories Earth-like, super-Earths/Neptunes, and giant planets. Overall, our distribution is compatible with the observed exoplanet populations, with the exception of the giant planets, whose observed distribution covers a wider parameter space than our simulation. The major difference we observe for cool stars is due to a strong observational bias: the majority of our simulated exoplanets orbit M-type stars, the most abundant stellar type (\citealt{Reyle2021} estimate that they represent around 60\% of all stars within 10 pc), which are underrepresented in the actual detections because they are so faint (for reference, they represent less than 2\% of the stars observed by \textit{Kepler}).

\subsection{How representative are exoplanet host stars of the underlying population?}

Comparing the properties of planet-hosting stars to those of the general stellar population is a common technique employed in observational exoplanet studies (e.g. \citealt{Adibekyan2012, Bashi2020, Tautvaisiene2022}). Indeed, these studies are necessary to calibrate the dependencies of the occurrence rate and multiplicity (see Sect. \ref{sec:occurrence}). 
Here, we compare the effect of our occurrence rate prescriptions on stellar parameters that do not directly influence the occurrence rates, such as age and birth radius (although both of these parameters are indirectly linked to metallicity through Galactic chemical evolution; see, e.g. \citealt{Chiappini2001, Minchev2018}).

\begin{figure}
    \centering
    \includegraphics[width=0.49\textwidth]{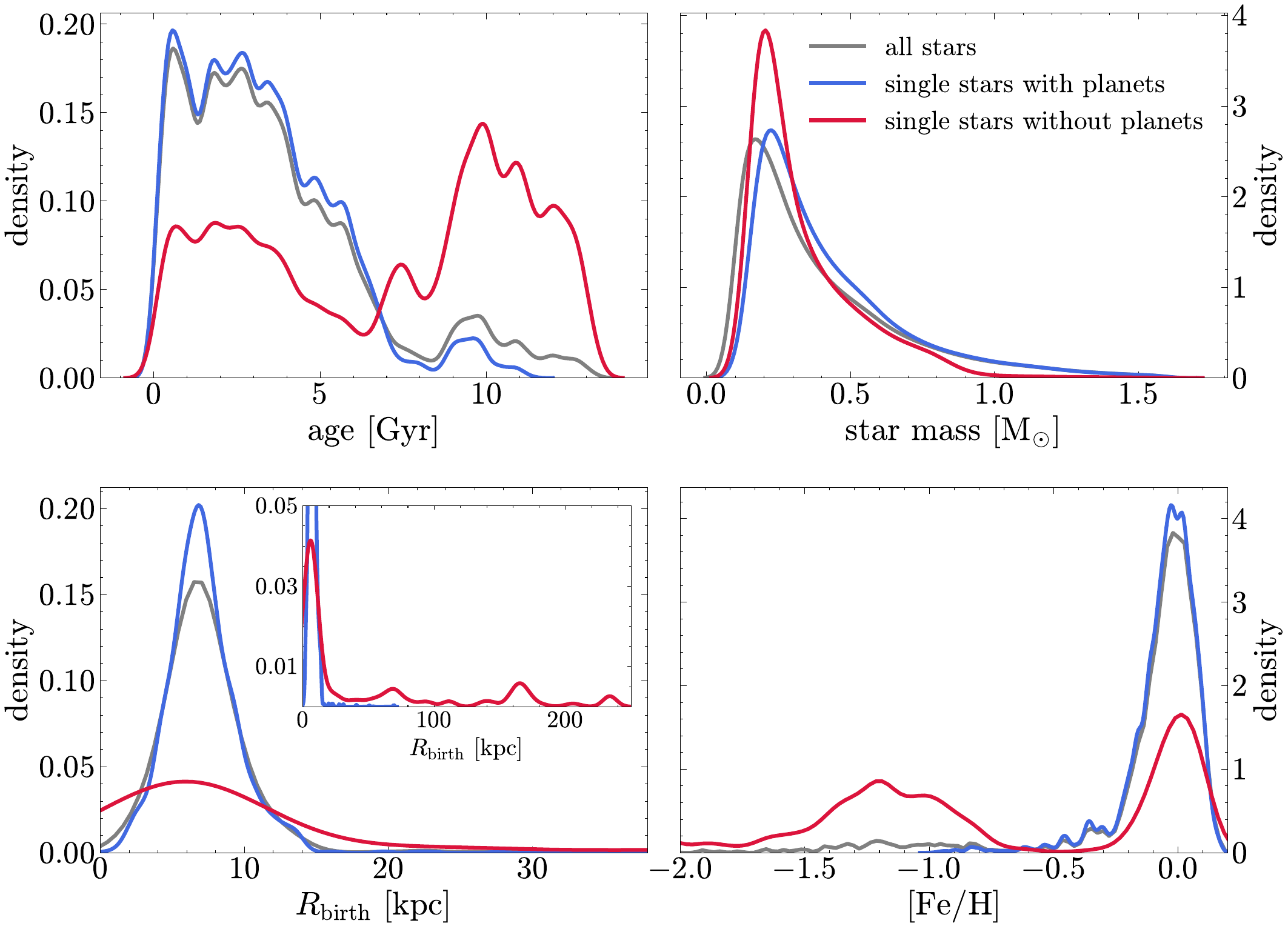}
    \caption{Stellar age, mass, birth radius, and metallicity density distributions for the SN simulation. In each panel, we show the distributions of all simulated stars, stars without planets, and stars with planets, as indicated in the legend (single stars without planets represent $\sim 14\%$ of the single star population).\label{fig:comparison_star_properties_with-out_planets}}
\end{figure}

Our model predicts that only 14\% of the single stars in the solar vicinity do not host exoplanets. In Fig.~\ref{fig:comparison_star_properties_with-out_planets} we compare the distribution in age, stellar mass, birth radius, and metallicity [Fe/H] of the entire population (``all stars'', in grey) with that of the single stars with planets (in blue) and single stars without planets (in red).
Since the fraction of planet-hosting stars is large, we find only subtle differences between the planet-hosting and the entire stellar population.
On the other hand, we do observe pronounced differences between non-hosting single stars and planet hosts. 
The most important difference is that around half of the non-hosting stellar population are metal-poor stars (and thus mostly old), while almost all hosting stars have a metallicity greater than $-0.5$. This is partly due to our modelling of the planet occurrence rate, set to 0 for stars with [Fe/H] $\le -1.0$; see Fig. \ref{fig:occ_rate_planets_mass_feh} and Sect. \ref{sec:occurrence}.
We also note that the non-hosting stars group contains a non-negligible fraction of accreted stars that were born outside of the main galaxy (see birth radius distribution on the lower left panel of Fig.~\ref{fig:comparison_star_properties_with-out_planets}). 

\subsection{Planets in the circumstellar habitable zone}\label{sec:habitability}

The circumstellar habitable zone (CHZ) is the region around a star in which liquid water might exist on the surface of orbiting planets. This zone is considered a key factor in determining the potential for life as we know it, since liquid water is essential for life on Earth. For this reason, one of the primary objectives of the {\it Kepler} mission was to determine the occurrence rate of rocky exoplanets in the CHZ of solar-like stars \citep{Borucki2010, Borucki2016}, often referred to as $\eta_{\oplus}$. Many studies have since derived estimates of $\eta_{\oplus}$, with values ranging from $<1\%$ to $\gtrsim 50\%$ (see, e.g. Appendix 2 of \citealt{Scherf2024}). The boundaries of the CHZ depend mainly on stellar luminosity and effective temperature, although other factors are sometimes included \citep[e.g.][]{Kaltenegger2017, Lammer2024}. 
The optimistic HZ is delimited empirically by the ``recent Venus'' and ``early Mars'' limits, i.e. set by the last time liquid surface water could have existed, respectively on Venus and Mars ($S/S_\oplus=1.78$ and $S/S_\oplus=0.32$ for $T_{\rm eff}=5800$ K; \citealp{Kasting1988, Kaltenegger2017}). 
Here, we use this optimistic definition of the CHZ (see Fig. \ref{fig:teff_instellation}).

In order to assess whether our simulated planets are in the CHZ of their host stars, we used the polynomial equations from \citet{Kopparapu2013} to determine the limits of the CHZ. 
We find that $\sim 23$\% of our solar-vicinity exoplanets lie in the CHZ of their host star ($\sim 25$\% of the Earth-likes, $\sim 20$\% of the super-Earths/Neptunes, and $\sim 29$\% of the giants), a result which is higher than most of the literature values cited above. This can be explained by the extreme under-representation of Earth-like planets in observations compared to the underlying population, as visible in Fig. \ref{fig:teff_instellation}.

\section{Forward-modelling the {\it Kepler} exoplanet census}\label{sec:kepler}

The significance of the {\it Kepler} mission \citet{Borucki2010} for exoplanetary science can hardly be overestimated \citep{Lissauer2023}. The {\it Kepler} field in the constellation of Cygnus, with its more than 4000 exoplanet candidates \citep{Lissauer2024}, is thus the natural testbench for exoplanet population models. In this section, we provide forward simulations of the exoplanetary population in the {\it Kepler} field based on the method described in Sects. \ref{sec:method} and \ref{sec:results_SN}, taking into account the most important selection effects of the original {\it Kepler} mission.
This allows us to evaluate the overall quality of our exoplanet simulation and helps us identify ways for future improvements.

\subsection{{\it Kepler} selection function \label{sec:selection_fct_kepler}}

\begin{figure}
    \centering
    \includegraphics[width=.49\textwidth]{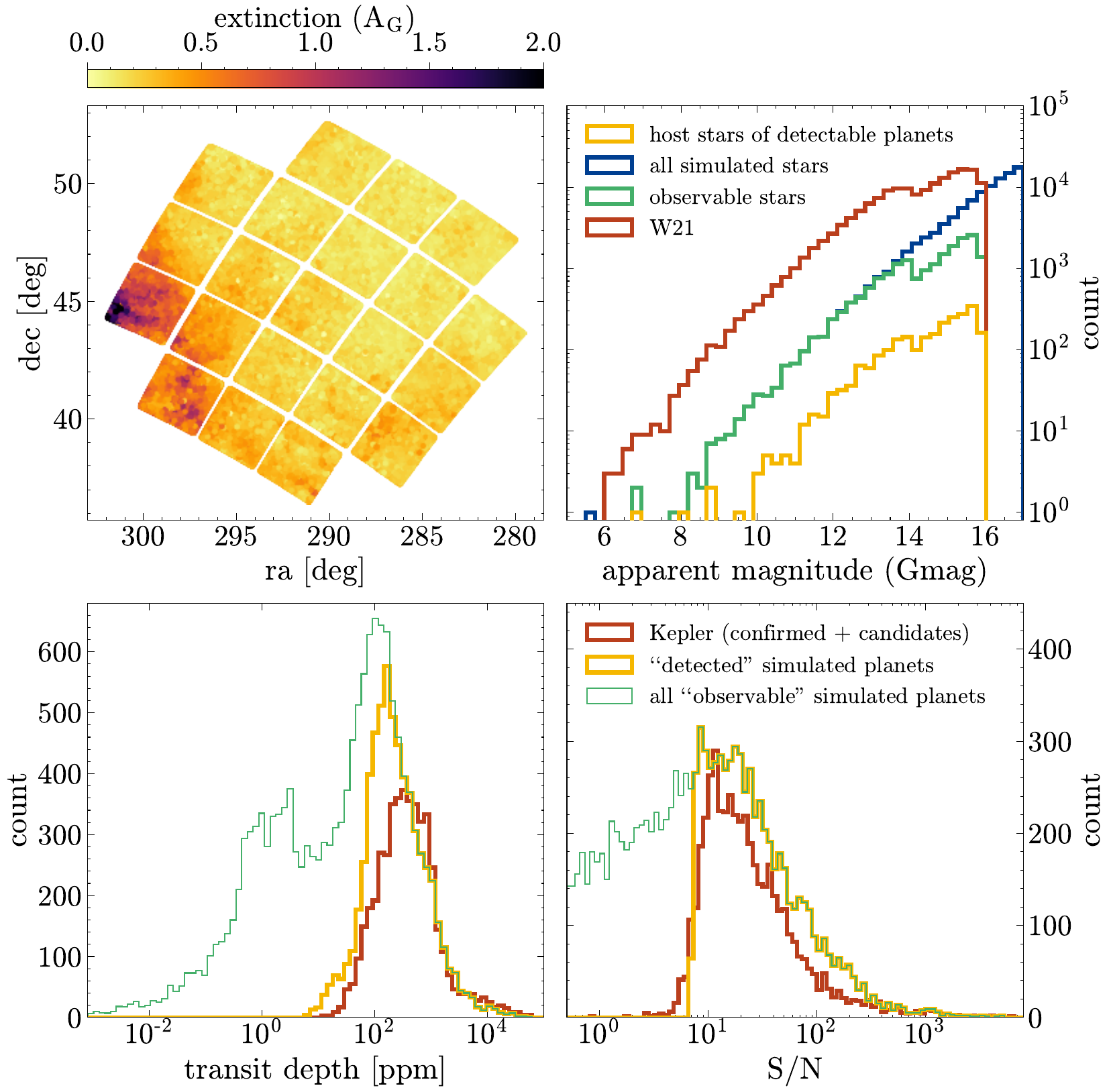}
    \caption{Summary of our simulation of the exoplanet population in the {\it Kepler} field. {\it Upper left panel}: geometric selection of stars in the simulated {\it Kepler} field, colour-coded by interstellar extinction in the $G$ band, $\rm A_G$. {\it Upper right}: Apparent $G$ magnitude distribution of all simulated stars in the field of view (in blue), the ``observable'' stars, as selected in Sect. \ref{sec:selection_fct_kepler} (green), compared to the stars observed by {\it Kepler}  (\citealt{Wolniewicz2021} catalogue, in red).
    {\it Lower panels}: Transit depth and transit signal-to-noise distributions of planets in the {\it Kepler} field. Red: observed distribution among candidate+confirmed {\it Kepler} exoplanets. Green: all simulated planets with orbit inclinations compatible with a transit observation (see Sect. \ref{sec:detectability}). Yellow: simulated detected exoplanets (i.e. planets with $\rm S/N \ge 7.1$).}
    \label{fig:kepler_field_hist_magn}
\end{figure}

To produce a simple mock sample of the {\it Kepler} field, we use the same galaxy {\tt g7.55e11} presented in Sect. \ref{sec:simulation}. We now focus on a sky area that matches the {\it Kepler} field in size and position (see Fig. \ref{fig:kepler_field_hist_magn} top left panel). Stars and planets are created following the methods described in Sect. \ref{sec:method}, with an additional dispersion in the galactic coordinates to ensure a smooth sampling: we first select all stellar particles in a wide cone around the \textit{Kepler} field of view up to a distance of 5.5 kpc from the Sun, to include all \textit{Kepler} exoplanets. Then, we generate the stellar population following the methods described in Sect. \ref{sec:stars}, before sampling the galactic coordinates of all created stars, following a Gaussian distribution centred on the coordinates of their host stellar particle, with a dispersion of 500 pc in $X_{\rm Gal}$ and $Y_{\rm Gal}$ and of 100 pc in $Z_{\rm Gal}$. Finally, we select the stars located in the \textit{Kepler} field, using \texttt{K2fov} package \citep{Mullally2016}.
Interstellar extinction, although low in most of the {\it Kepler} field \citep{Rodrigues2014}, was added using the 3D extinction map of \citet{Green2019} included in the {\tt dustmaps} package \citep{Green2018}. Finally, we converted the obtained reddening $E(B-V)$ to extinction in the {\it Gaia} $G$ band using $A_G \simeq 0.84\cdot A_V$, and $A_V = 3.1\cdot E(B-V)$.
The top left panel of Fig. \ref{fig:kepler_field_hist_magn} shows the resulting extinction distribution.

To obtain a stellar population comparable to the \textit{Kepler} target list \citep{Batalha2010}, we apply the selection fraction as a function of magnitude derived by
\citet{Wolniewicz2021} (their Fig. 2b) to the full sample of generated stars. Additionally, we discard all stars with apparent magnitude $G > 16$, as well as giants (defined as $\log(g) \le 4$) with magnitudes $G > 14$, to match the actual observed fraction (their Fig. 2a). 

After those two magnitude cuts, our sample contains $\sim 31\,000$ stars, with an overrepresentation of stars with magnitudes between $G=15.5$ and $G=16$. The target selection of \textit{Kepler} is based on criteria we did not consider, such as the minimal planetary radius expected to be detectable, $R_{\rm p, min}$ \citep{Batalha2010}. As we are not able to compute this quantity for our simulated population, we modified the magnitude selection function taken from \citet{Wolniewicz2021} to match the final magnitude distribution of {\it Kepler} target stars \citep{Wolniewicz2021, Farmer2013} (top right panel of Fig. \ref{fig:kepler_field_hist_magn}). We end up with around $20\,000$ ``observed'' stars in the \textit{Kepler} field, this number being $\sim 12\%$ of the stars actually observed by \textit{Kepler}\footnote{Using the \cite{Wolniewicz2021} catalogue with the \texttt{Observed} flag of 1, and applying cuts in magnitude ($\in [6, 16]$) and mass ($\le 1.6 ~M_{\oplus}$) to be comparable to our stellar population. Those additional cuts have a minimal impact on the total number of stars.}. One of the factors explaining this difference is that we consider only the single stars from our simulation, while some observed stars in the \textit{Kepler} field are unresolved binaries. However, including all the primary stars (making the same magnitude cuts) would not even double our observed star population. The difference in star counts is thus mainly due to the lower stellar density in the simulation compared with the actual SN (see the first paragraph of Sect. \ref{sec:results_SN}).

\begin{figure*}
    \centering
    \includegraphics[width=1.\textwidth]{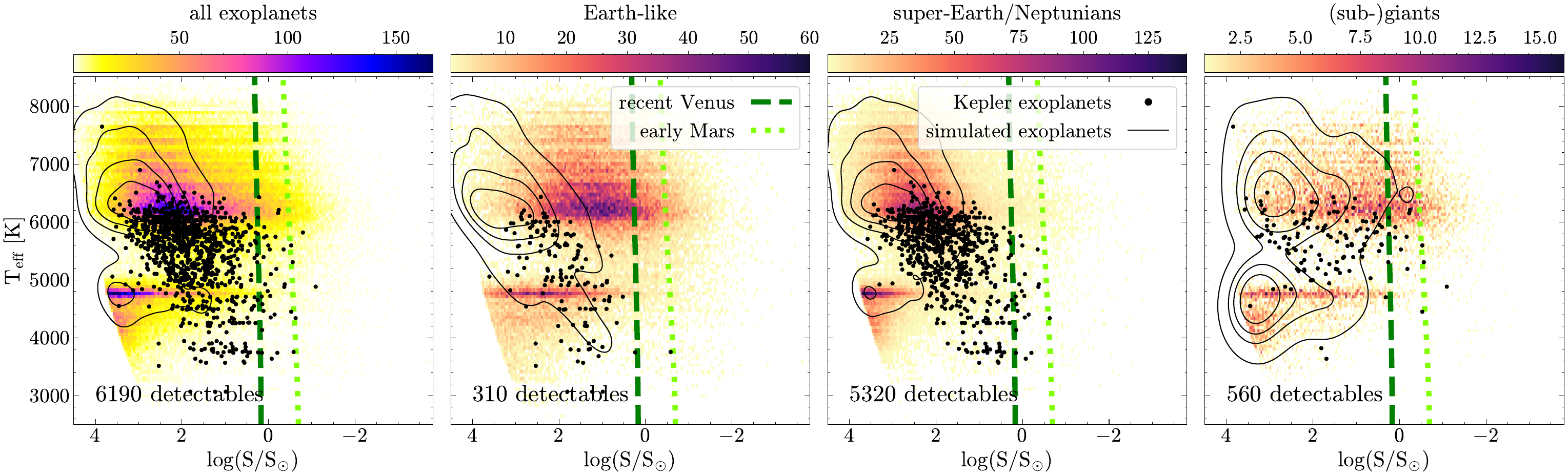}
    \caption{Effective temperature vs. instellation, analogous to Fig. \ref{fig:teff_instellation}, but for the simulated \textit{Kepler} field. The distribution of the exoplanets observed by {\it Kepler} is shown as black dots, while the distribution of ``detectable'' exoplanets in the {\it Kepler} simulation is shown in black density contours. The background distribution shows the distribution of 
    planets after applying magnitude cuts described in Sect.~\ref{sec:selection_fct_kepler}. The number of detectable exoplanets in the simulated \textit{Kepler} field is indicated on the lower left part of each panel.
    }
    \label{fig:Teff_insolation_simu-detected_Kepler}
\end{figure*}

\subsection{Selection of observable transiting exoplanets} \label{sec:detectability}

We apply our exoplanet generation process to the simulated stars which would be ``observed'' by \textit{Kepler} (around $20\,000$, see Sect. \ref{sec:selection_fct_kepler}), and obtain around $170\,000$ exoplanets.
Assuming a random inclination (see Appendix \ref{annexe:CDPP}) and ephemeris distribution of the planets orbiting each star, we calculate the detection probability of each exoplanet during a {\it Kepler}-like single-star observation campaign of 4 years.
In order to simulate the observability of synthetic exoplanets by \textit{Kepler}, we first estimate the signal-to-noise ratio (S/N) of each exoplanet transit. In practice, we use equation 4 of \cite{Christiansen2012}: 
\begin{equation} \label{eq:SNR_christiansen12}
    {\rm S/N} = \sqrt{N_{\rm tr}} \times \frac{\delta}{\rm CDPP_{\rm eff}} \text{, with }  {N_{\rm tr}} = \frac{{t_{\rm obs}} \times f_{\rm o}}{P}.
\end{equation}

In this equation, the number of observed transits $N_{\rm tr}$ is obtained from the planet's orbital period $P$, the total observational time of \textit{Kepler} $t_{\rm obs}$,  and the fraction of this total time when the target was observed, $f_{\rm o} = 0.92$, following \cite{Christiansen2012}.
The transit depth $\delta = (R_{\rm pl} / R_{\rm st})^2$ is the squared ratio of the planet's and host star's radius (shown in the lower left panel of Fig. \ref{fig:kepler_field_hist_magn}). Planet radii $R_{\rm pl}$ are estimated from the planet masses, see Appendix \ref{sec:annexe_mass_to_rad} for details. 
The secondary peak around 1 ppm in the transit-depth distribution is due to the (mostly undetectable) Earth-like population.

Finally, the $\rm CDPP_{eff}$ is the effective combined differential photometric precision, a photometric precision metric. The $\rm CDPP_{eff}$ was experimentally measured during the \textit{Kepler} mission \citep{koch_kepler_2010, Jenkins2010}; we give more details on how we estimate it for our synthetic planets in Appendix \ref{annexe:CDPP}.
Finally, we consider a planet to be synthetically ``detected'' by \textit{Kepler} if its $\rm S/N \ge 7.1$: our fiducial ``detected'' sample for the \textit{Kepler} field of view contains around 6200 planets, orbiting 2400 host stars\footnote{Mean value among 1000 iterations - These numbers can change slightly ($\pm154$ planets, $\pm46$ host stars) with each random rerun. The number of planets per star does not vary significantly ($2.56\pm0.04$).}. The final S/N distribution is shown in the bottom right panel of Fig. \ref{fig:kepler_field_hist_magn}.

To test our synthetic S/N estimates on real data, we apply them to all candidates and confirmed planets from {\it Kepler}. The results are shown in Fig. \ref{fig:comparaison_SNR_Kepler}: the one-to-one comparisons with {\it Kepler} demonstrate that 1. the transit depth is retrieved with very good accuracy and precision, and 2. our S/N estimates are globally accurate, albeit with a slightly higher dispersion compared to the published values by \textit{Kepler} (\texttt{koi\_model\_snr}), see Appendix~\ref{sec:annexe_trdepth_snr_comparison_kepler}.

\subsection{Comparison of simulated ``detectable'' exoplanets with {\it Kepler} planets and candidates}

We now proceed to a more detailed comparison of the {\it Kepler} exoplanet census with our tailored simulation described in the previous subsections. This allows us to identify potential problems with our simulations and to propose ways to overcome them in the future.

In the lower panels of Fig. \ref{fig:kepler_field_hist_magn}, we compare the transit depth and S/N distributions of our simulated ``detected'' exoplanet sample with the {\it Kepler} exoplanet catalogue. We obtain similar distributions, with an overdensity of planets with transit depths between $10$ and $300$ compared to the {\it Kepler} data. 
The total number of ``detected'' planets (around 6200 planets around 2400 host stars) is similar to the observed population ({\it Kepler} DR25: 4619 around 3024 host stars), despite having $\sim 10$ times fewer simulated stars than observed ones. We notice that we tend to observe too many exoplanets per star (see Sect. \ref{sec:keplerFOV_occrate}).
This is likely a consequence of the fact that 1. we do not model stellar variability which further degrades the CDPP, 2. we assume that all planets of a system are perfectly aligned in the same orbital plane, and 3. we estimate each exoplanet's detectability individually, even in multi-planetary systems (a minor effect for transit surveys like {\it Kepler}). 

\subsubsection{Comparison of host star properties}

Starting with the {\it Kepler} Input Catalogue \citep{Brown2011}, several works have presented stellar parameters for stars (and in particular planet hosts) in the {\it Kepler} field (e.g. \citealt{Huber2014, Johnson2017, Berger2020, Zhang2025}). For our comparison, we use the homogeneous stellar parameters obtained from {\it Gaia} DR3 low-resolution XP spectra and parallaxes combined with broad-band photometry by \citet{Khalatyan2024}.

In Appendix \ref{annexe:comparison_stellarparams_hoststars_Kep_fov}, we compare the distributions of apparent magnitudes, distances, effective temperatures, and stellar radii of our synthetic ``detected'' host stars with the ones from the {\it Kepler} catalogue (candidates + confirmed exoplanet hosts). 
The key differences are: 1. Our simulation predicts a larger population of F-type main-sequence planet hosts than observed, and 2. our simulation also predicts more giant-star planet hosts than observed by {\it Kepler} (see bottom panels of Fig. \ref{fig:comparison_stellarparams_hoststars_Kep_fov}). These discrepancies arise because the {\tt g7.55e11} simulation differs from the Milky Way disc in some important parameters. Among them, one notable difference is the over-representation of young stars, caused by a rather late peak in the star-formation history, and leading to an over-representation of hot stars ($\rm T_{eff} > 6000~K$, see middle left panel of Fig. \ref{fig:comparison_stellarparams_hoststars_Kep_fov}).
The distance distribution discrepancy (top right panel) can be explained by the larger vertical velocity dispersion and scale height of the disc in \texttt{g7.55e11} compared to the MW (see also Sect. \ref{sec:caveats}).

\subsubsection{Comparison of planet occurrence rates}
\label{sec:keplerFOV_occrate}
As discussed in Sect. \ref{sec:detectability}, although the number of our simulated ``detectable'' exoplanets in the {\it Kepler} field is comparable to the confirmed+candidate exoplanets detected by {\it Kepler} (several thousands), we find too many planets per star, with a mean of $2.56 \pm 0.04$ planets per host star, versus $\sim 1.3$ for \textit{Kepler}. In Fig. \ref{fig:Teff_insolation_simu-detected_Kepler}, we show a comparison of the observed and simulated exoplanet families in the $T_{\rm eff}$ vs. instellation plane (analogous to Fig. \ref{fig:teff_instellation}). We see that our simulation overpredicts the number of SE/Neptunians, while it underpredicts the number of Earth-like planets: we obtain 5\% of Earth-like planets, 85\% of SE/Neptunes, and 10\% of giants, while \textit{Kepler} confirmed exoplanets are composed of 14\% of Earth-like, 74\% of SE/Neptunes and 12\% of giants.

The percentage obtained for each planet type depends on the definition we adopt. While the percentage of Earth-like planets is stable throughout the different definitions we compared, the percentage of SE/Neptunes and sub-giants/giants varies more strongly. As discussed in Sect. \ref{sec:mass-period}, we choose to use the mass-based definition for our simulated exoplanets in Sects. \ref{sec:results_SN}, \ref{sec:same_glx_different_regions}, and \ref{sec:different_glx_same_region}. For our comparison with the \textit{Kepler} planets in this section, however, we chose to use the radius-based definition, since very few of the {\it Kepler} planets have mass estimates.

\subsubsection{Comparison of planet masses and radii distributions}

\begin{figure}
    \centering
    \includegraphics[width=.49\textwidth]{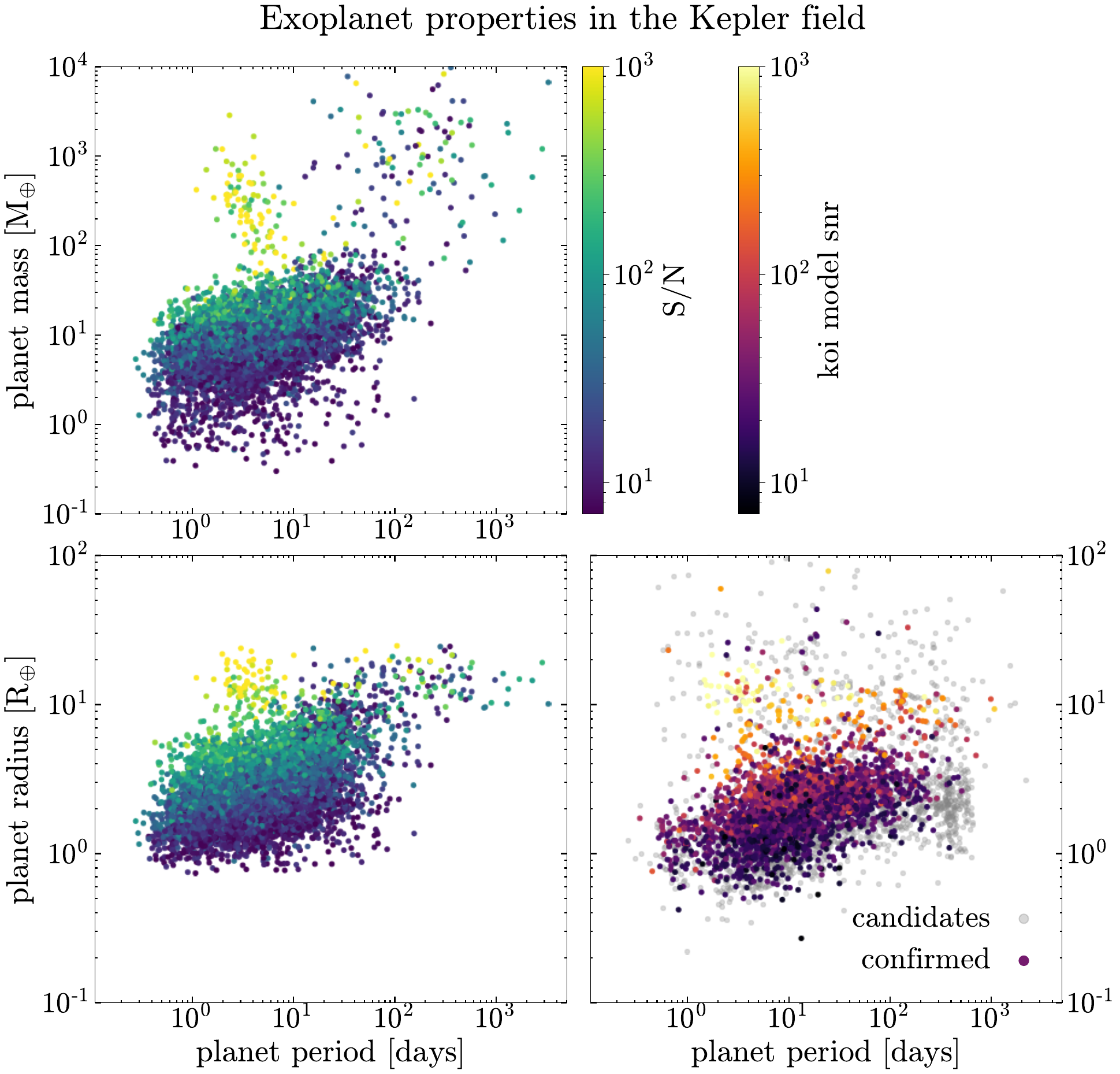}
    \caption{Comparison with exoplanets in the \textit{Kepler} field (mass-period and radius-period diagrams). The left column shows the exoplanet population obtained by our simulation. The right plot shows the exoplanets found by \textit{Kepler} (both candidates and confirmed). Top panel: planet mass (in $\rm M_{\oplus}$) vs. orbital period (in days). Bottom row: planet radius (in $\rm R_{\oplus}$) vs. period. For the simulated exoplanets, the radius was obtained with a conversion from the planet mass (see Appendix \ref{sec:annexe_kepler}).}
\label{fig:kepler_period_radius}
\end{figure}

Fig. \ref{fig:kepler_period_radius} shows the mass-period and the period-radius diagrams of our forward-simulated ``detected'' exoplanets, compared with the observed {\it Kepler} population. The mass-period diagram follows, by construction, the populations shown in Fig. \ref{fig:MP_fit_Drazkowska}, modulo the (heavy) selection effects simulated in Sects. \ref{sec:selection_fct_kepler} and \ref{sec:detectability}. Since precise masses are not readily available for most of the observed {\it Kepler} exoplanets, we only show the observed population in the period-radius diagram (lower panels of Fig. \ref{fig:kepler_period_radius}).

After simulating detectability, we have fewer planets with small radii than the actual confirmed cases by \textit{Kepler}. We suggest that this discrepancy is caused by 1. planets that were confirmed by follow-up observations, and 2. our mass-to-radius conversion (see Sect. \ref{sec:annexe_mass_to_rad}), which might be too simplistic.
Another well-known feature of the {\it Kepler} data that we do not reproduce is the so-called ``radius gap'', a bimodality in the radius distribution of small planets \citep[e.g.][]{Fulton2017, Zeng2017, Petigura2022}, which is expected since we did not explicitly include it in our modelling (see caveats in Sect. \ref{sec:caveats}). 

Another subtle effect that is seen in the data but not in our model is the photo-evaporation of close-in planets (with periods $\lesssim 3$ days): planets that initially lay between the low-density sub-Neptunes and the dense super-Earths are likely to lose their atmospheres by the intense radiation of the nearby host star, which results in a gradual decrease of the planet radius over time (see \citealt{McDonald2019} for a summary). They end up in lower regions of the $R-P$ space. This phenomenon is complex, depending on the planet's and atmosphere's composition, as well as on the stellar type of its host star. We therefore confirm that neither the radius gap nor the missing close-in Neptunes (generally attributed to photo-evaporation) are caused by selection effects. In the future, we aim to introduce some of these more intricate properties of the exoplanet population into our modelling.

\section{Exoplanet populations in different regions of the simulated galaxy} \label{sec:same_glx_different_regions}

The impact of the galactic environment on planet formation and evolution has been debated since the seminal paper of \citet{gonzalez_galactic_2001} who coined the term Galactic Habitable Zone (see also e.g. \citealt{Lineweaver2004, Prantzos2008, Gowanlock2011, Baba2024}). In this and the following section, we refrain from discussing this concept in detail, but study the distribution of different types of exoplanets as a function of galactic environment, using the same simulation framework used to study the solar vicinity (Sect. \ref{sec:results_SN}) and the {\it Kepler} field (Sect. \ref{sec:kepler}). 
\citet{Bashi2022} opened the studies about the influence of the Galactic context on the exoplanet formation, with a sample of 506 planet candidates from \textit{Kepler} DR25 orbiting 369 host stars, associated to either the thin disc (292 stars), the thick disc (7 stars) or the halo (0 star). They found that thin disc stars (mostly young and metal-rich) host on average more planets than thick disc stars (mostly old and metal-poor). Their sample being limited by its size, it is interesting to use Galactic simulations to study the exoplanet distribution in the Galaxy.

\begin{figure*}
    \centering
    \includegraphics[width=1\textwidth]{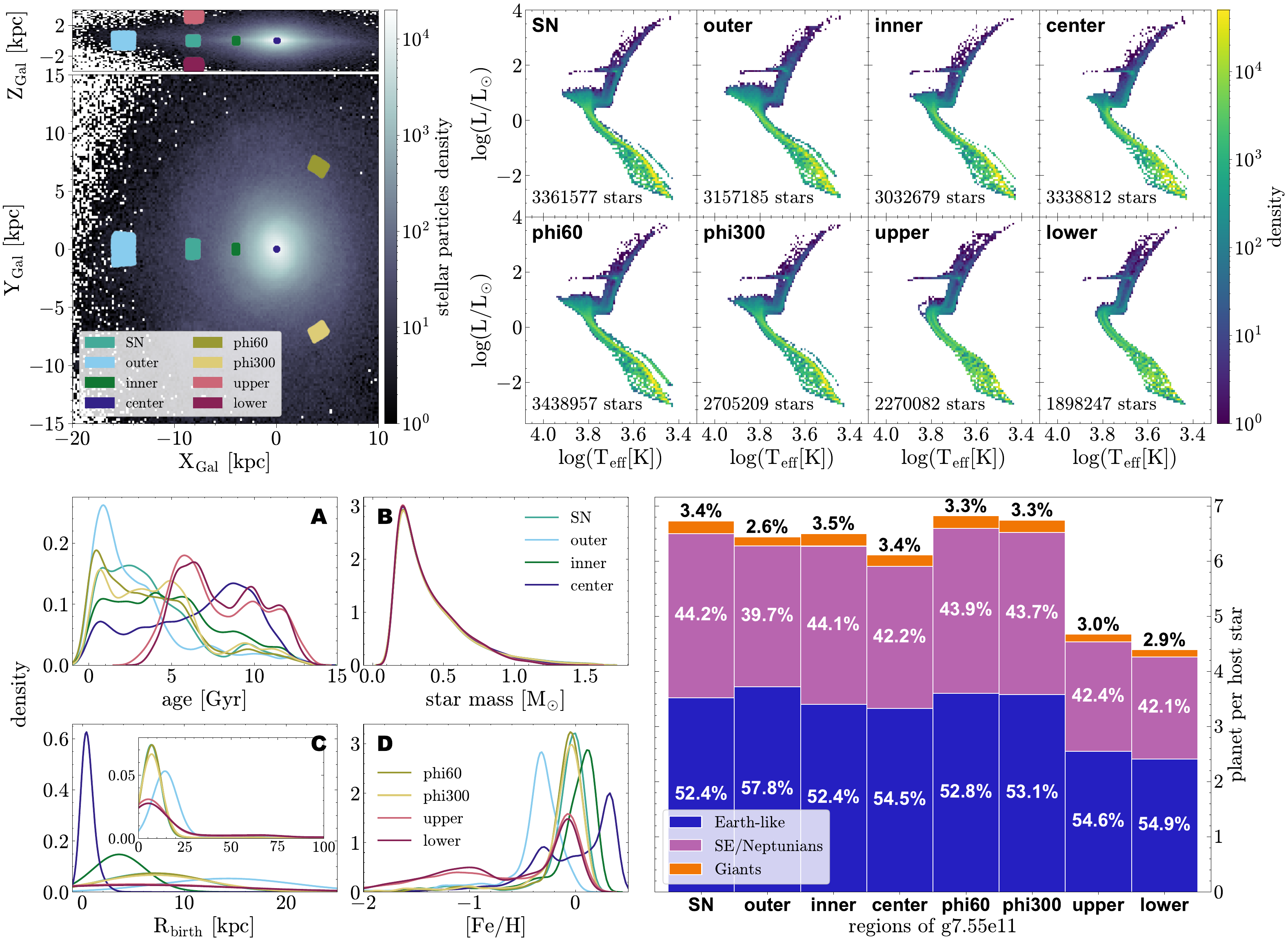}
    \caption{Results of our planetary population synthesis for different regions in the MW-like galaxy g7.55e11. Top left panels: Illustration of the regions studied, in Galactocentric Cartesian co-ordinates. Top right panels: metallicity-coloured Hertzsprung-Russell diagrams for each of the regions. Bottom left panels: Stellar age (A), stellar mass (B), birth radius (C), and metallicity (D) distributions, analogous to Fig. \ref{fig:comparison_star_properties_with-out_planets}. Bottom right panel: Relative planet occurrence statistics for each of the studied regions.
    }
\label{fig:g755_regions_summary}
\end{figure*}

In this section, we study the simulated exoplanet populations in eight regions of the galaxy {\tt g7.55e11} (summarised in Fig. \ref{fig:g755_regions_summary}).
In addition to the solar vicinity sample studied in Sect. \ref{sec:results_SN}, we selected seven regions that cover the full extent of the galaxy's disc and bulge, each containing a similar number ($\sim 1000$) of stellar particles. The top left panel of Fig.~\ref{fig:g755_regions_summary} shows the location of the seven additional regions: i) an outer-disc region  ($R_{\rm Gal}\sim 15$ kpc; cyan), ii) an inner-disc region ($R_{\rm Gal}\sim 4$ kpc; green), iii) the galactic centre (violet), iv) two regions similar to the SN selection but at different angles ($\phi_{\rm Gal} = 60^{\circ}, 300^{\circ}$, respectively; yellowish colours); v) two regions with the same distance to the galactic centre than the SN, but situated above and below (2.5 kpc $< \lvert Z_{\rm Gal} \rvert < 4$ kpc) the galactic disc, respectively (red colours).

In the top right plots of Fig.~\ref{fig:g755_regions_summary}, we show the Hertzsprung-Russell diagrams of the simulated stars in each of the eight regions highlighted in the top left panel of that figure, reflecting the impact of the different age and metallicity distributions on the stellar population in each region. 
The bottom row of Fig. \ref{fig:g755_regions_summary} shows the distribution of the simulated stellar population in age, mass, birth radius, and metallicity (lower left panels) and the relative planet occurrence statistics (lower right panels) for each of the eight selected regions. 

We first analyse the stellar populations of the eight regions. We observe very little variation in the stellar mass distribution -- this is expected because we assume the IMF to be invariant. Regarding the metallicity distributions within the different regions, we find a clear signature of the negative radial metallicity gradient: the galactic centre region is predominantly metal-rich, and the peak [Fe/H] moves to gradually lower values for the \texttt{inner} region, the \texttt{SN}, \texttt{phi60}, \texttt{phi300}, the \texttt{upper} and \texttt{lower}, and finally the \texttt{outer} region. The \texttt{upper} and \texttt{lower} regions both contain a significant metal-poor tail, as expected, corresponding to potential accreted/halo stars.
Studying the host star population in those metal-poor regions is interesting to verify whether such objects can be observationally found and to further explore those regions in search of potential accreted exoplanetary systems, as proposed by \citet{Perottoni2021}.
The \texttt{centre} region has a secondary peak around $\rm [Fe/H]\simeq-0.4$, reflecting the complex stellar population mixture that we also observe in the centre of the Milky Way (e.g. \citealt{Schultheis2015, Schultheis2020, Nieuwmunster2023, Nandakumar2024}). 

The birth radius distributions of all regions are approximately Gaussian, centred on the central radius of the respective region. The dispersion in birth radius increases with distance from the galactic centre (and distance from the disc plane, for the high-$|Z|$ regions). This is mostly due to an increase in the fraction of accreted stars towards the outer disc, while radial migration within the disc dominates in the inner/SN regions.
Finally, the age distributions in Fig. \ref{fig:g755_regions_summary} reflect the significant differences among the star-formation histories of the eight regions of {\tt g7.55e11}.  As expected, the two high-$|Z|$ regions do not contain any young stars ($< 4$ Gyr). The age distributions reflect the inside-out formation of the galaxy, with the peak of the age distribution shifting towards lower ages as we move from the inner to the outer galaxy. Consistent with its metallicity distribution, the age distribution of the galactic-centre sample is also very broad.

Regarding the exoplanet populations in the eight regions of {\tt g7.55e11}, we find that the number of planets per star is similar between all regions (between 6 and 7), except for the \texttt{upper} and \texttt{lower} regions which present a mean number of 4 planets per star (due to the lower mean metallicity of the stars found in those regions). The planet type fraction is similar in all of them ($52-58$\% Earth-like, $39-44$\% SE/Neptunes, $2.6-3.5$\% giant planets), with slight variations caused by the stellar-population differences. For example, the \texttt{outer} region displays a relative fraction of Earth-like planets that is $\sim 5\%$ higher than in other regions, and the lowest percentage of giant planets (see Sect. \ref{sec:discussion} for a discussion).

Recently, \citet{Boettner2024} also investigated variations in galactic planet populations using a different methodology. They combined a high-resolution Milky Way analogue ("\texttt{37\_11} run") from the HESTIA suite of simulations \citep{Libeskind2020} with the Bern planet formation model (the New Generation Planet Population Synthesis, NGPPS; \citealt{Emsenhuber2021b}). 
The stellar particles of this galactic snapshot have a typical mass of $\sim 10^4 \rm ~M_{\odot}$. They created the stellar population from all the stellar particles, assuming uniform age and metallicity, and sampling masses from the \cite{Chabrier2003} IMF around a unique value $M_{\star} = 1~\rm M_{\odot} \pm 5\%$, keeping only the main sequence stars, to obtain a total number of stars $N_{\star}$. For each stellar particle, they assign a certain number $n$ of planets of different categories per host star, obtained from the NGPPS model (assuming the initial number of embryos $N_{embryos} = 50$ and the stellar mass $M_{\star} = 1~\rm M_{\odot}$).
They finally obtain the total number of planets: $N_{pl} = n \times N_{\star}$.
Applying this framework, they compared the obtained occurrence rates in metal-rich regions (bulge, thin disc) with metal-poor regions (thick disc, halo). Discrepant from our findings (in which the relative proportions among different planet types remain fairly constant throughout the Galaxy; see Fig. \ref{fig:g755_regions_summary}), they found that giant planets ($>300~M_{\oplus}$) were $10-20$ times more frequent in the thin disc, while Earth-like planets were $1.5$ times more abundant in the thick disc. Low-mass planets were widespread across all regions. Comparing yields using uniform-mass population of lower stellar mass ($0.3-0.5 ~\rm M_{\odot}$), they found that the trends were weaker for low-mass dwarfs than solar-mass stars, leading to more homogeneous results across the galaxy. As they base their galactic analysis only on $1~\rm M_{\odot}$ stars, while we consider a broader range of stellar masses, we expected to obtain weaker occurrence variation between galactic environment.

\section{Comparison of the same region in different simulated galaxies} \label{sec:different_glx_same_region}

In this section, we investigate the impact of the simulated galaxy choice on our results. It is well known that the morphology and physical properties of disc galaxies vary considerably \citep[e.g.]{Dutton2010, Dutton2017}. Even when ``Milky-Way-like'' galaxies are explicitly selected from large simulation suites (e.g. by mass, rotation curve, bar, and spiral features), we often find striking differences between the individual galaxies \citep[e.g.][]{Grand2024, Pillepich2024}. We therefore investigate how the details in the evolutionary history of a galaxy impact the resulting exoplanet population in a solar neighbourhood-like volume, using the other five publicly available final snapshots from the NIHAO-UHD simulation suite of \citet{Buck2020}.

\begin{figure*}
    \centering
\includegraphics[width=1.\textwidth]{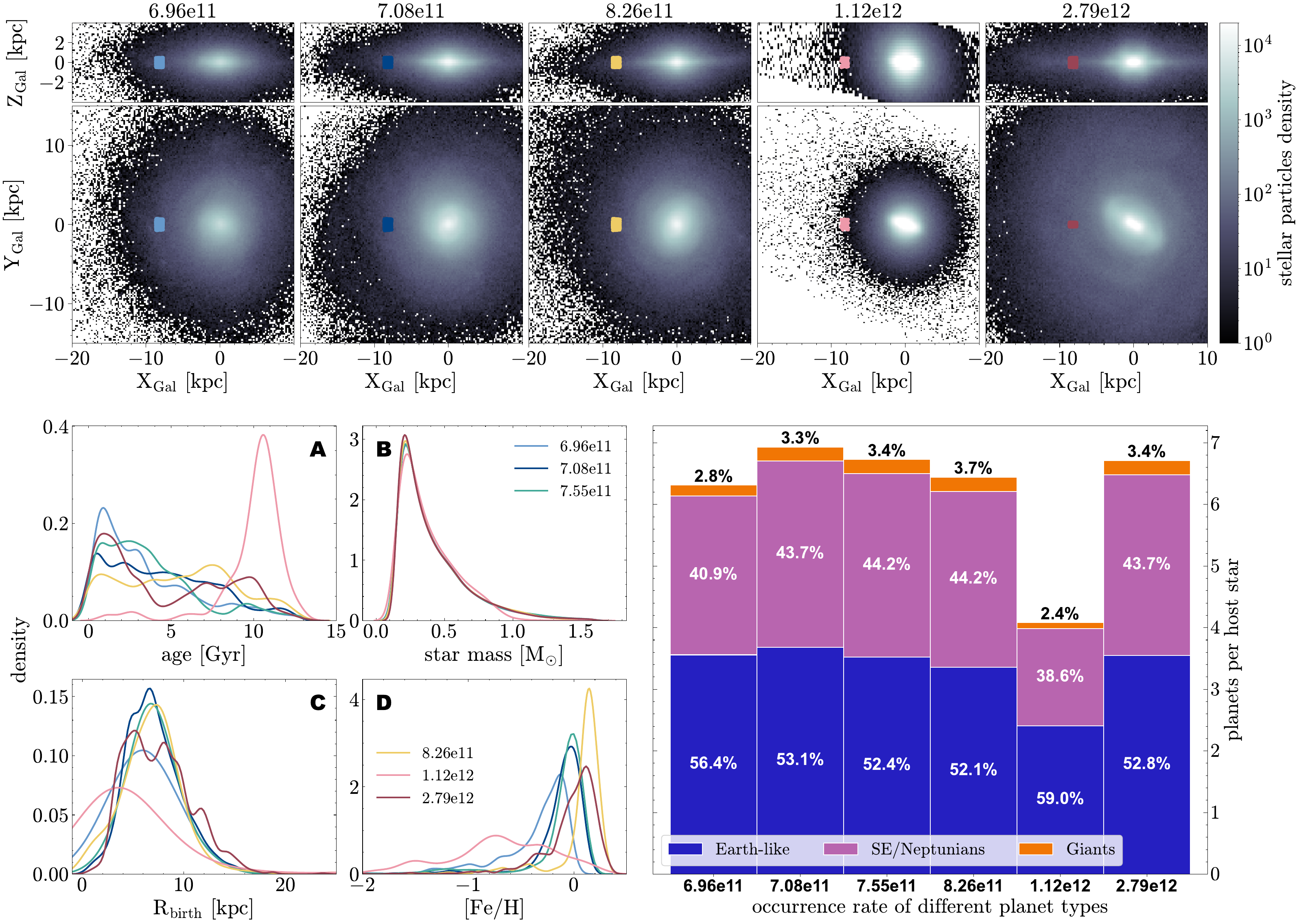}
\caption{Results of our planetary population synthesis for solar neighbourhood analogues in the six publicly available NIHAO-UHD galaxies, similar to Fig. \ref{fig:g755_regions_summary}. The top row highlights the selections of the solar vicinity volumes (8.2 kpc from the galactic centre, see Sect. \ref{sec:simulation}) in each of the final snapshots. Each selection, except for the more lenticular galaxy \texttt{1.12e12}, contains between 2 and 4 million stars. The lower panels show the simulated stellar and exoplanet statistics and are analogous to Fig. \ref{fig:g755_regions_summary}.}
\label{fig:6galaxies_summary}
\end{figure*}

In order to compare the solar-neighbourhood-like populations of different galaxies, we use the same spatial definition used in Sect. \ref{sec:simulation} for all final snapshots (see Fig. \ref{fig:6galaxies_summary}, top panel)\footnote{Because of its high mass and density in the selected region, we made an exception for the \texttt{g2.79e12} galaxy, reducing the angle of selection to [178, 182]$^\circ$ to keep a reasonable number of stellar particles.}. The differences between the six simulated galaxies are extensively discussed in \citet{Buck2020}; we summarised their main characteristics in Table~\ref{tab:nihao_simu_properties}.
All except \texttt{g1.12e12}, which has a lenticular shape, are clearly disc galaxies with spiral features. The most massive system, \texttt{g2.79e12}, also exhibits a prominent bar.

\begin{table}[h]
\caption{Main properties of the 6 simulated galaxies from NIHAO-UHD suite, compared to the MW.}
\label{tab:nihao_simu_properties}
\centering
\resizebox{\columnwidth}{!}{
\begin{tabular}{c c c c c c}
\hline\hline
    name & $M_{\rm star}$ & $R_d$ & $h_z^{t}$ & $h_z^{T}$ & $\langle M_{\rm st. part} \rangle$ \\
         & ($10^{10}~\rm M_{\odot}$) & (kpc) & (kpc) & (kpc) & ($10^3 ~\rm M_{\odot}$) \\
    \hline
    g6.96e11 & $1.58$ & 5.70 & --- & 1.4 & 5.8 \\
    g7.08e11 & $2.00$ & 3.90 & --- & 1.0 & 4.1 \\
    g7.55e11 & $2.72$ & 4.41 & 0.4 & 1.4 & 5.7 \\
    g8.26e11 & $3.40$ & 5.12 & 0.4 & 1.4 & 8.1 \\
    g1.12e12 & $6.32$ & 5.69 & --- & --- & 5.7 \\
    g2.79e12 & $15.9$ & 5.57 & 0.3 & 1.3 & 19.3 \\ 
    \multirow{2}{*}{MW}  & $2.6 \pm 0.4$\tablefootmark{a}; & \multirow{2}{*}{$\sim 2.6 \pm 0.5$\tablefootmark{b}} & \multirow{2}{*}{$\sim 0.22-0.45$\tablefootmark{b}} & \multirow{2}{*}{$\sim 1.45$\tablefootmark{b}} & \multirow{2}{*}{---} \\
        & $5 \pm 1$\tablefootmark{b} & & & & \\
    \hline
\end{tabular}
}
\tablefoot{$M_{\rm star}$ is the total stellar mass, $R_d$ the disc scale length, $h_z^t$ and $h_z^t$ the scale height at solar radius ($8 \pm 1$ kpc) of respectively the thin and thick discs, and $\langle M_{\rm st. part} \rangle$ the mean mass of the stellar particles. All parameters of the NIHAO-UHD galaxies are from \citet{Buck2020}. For MW values:
\tablefoottext{a}{\citet{Lian2025}}
\tablefoottext{b}{\citet{Bland-Hawthorn2016}}
}
\end{table}

The lower part of Fig.~\ref{fig:6galaxies_summary} shows the distribution of the simulated stellar population in age, mass, birth radius, and metallicity (lower-left panels) and the relative planet occurrence statistics (lower-right panel) for each of the simulated solar neighbourhoods in the NIHAO-UHD simulation suite.

Fig. \ref{fig:6galaxies_summary} demonstrates that the exoplanet population of a simulated solar neighbourhood does not depend drastically on the particular galaxy and its past history (reflected by the age, birth radius, and metallicity distributions), as long as it is broadly compatible with the main features of the Milky Way. This is illustrated by the little variations we see in the relative proportions of exoplanet types (lower right panel of Fig. \ref{fig:6galaxies_summary}). 
Even though there are important differences in the star-formation histories of the six galaxies, the only galaxies for which the relative occurrence rates of Earth-like vs. gaseous planets differs significantly from our fiducial simulation ({\tt g7.55e11} discussed above) are {\tt 1.12e12} (the lenticular galaxy, which hosts an important old and metal-poor population) and, to a lesser extent, {\tt 6.96e11}. This galaxy is smaller than our fiducial galaxy, and the chosen solar vicinity is therefore relatively further outwards from its galactic centre, as well as metal-poorer (and thus compatible with our findings in Sect. \ref{sec:same_glx_different_regions}). 

\section{Discussion and conclusions}\label{sec:discussion}

In this paper, we have studied the exoplanet population from a galactic perspective. We have created a framework that generates a synthetic exoplanet population based on a state-of-the-art galaxy simulation that reproduces the most important features of the Milky Way. Focussing first on the Solar Neighbourhood (SN; Sect. \ref{sec:results_SN}), we construct stellar populations and allocate planetary systems to single stars using a prescription informed by both exoplanet statistics and planetary formation models.

For the $3.4$ million single stars generated in the SN, we obtain $22.6$ million planets. They are distributed in $2.9$ million planetary systems, with a mean of $6.7$ exoplanets per star ($\sim 7.8$ per host star). The minority of single stars that do not host any planets are globally older and more metal-poor than the general population (Fig. \ref{fig:comparison_star_properties_with-out_planets}). Using mass-based planet type definitions, $52.5\%$ of the generated planets are classified as Earth-like, $44\%$ as SE/Neptunes, and $3.5\%$ as sub-giants/giants. Among them, $22.5\%$ are located in their host star's CHZ ($24.5\%$ of the Earth-like, $19.8\%$ of the SE/Neptunes, and $29.1\%$ of the sub-giants/giants; Fig. \ref{fig:teff_instellation}). Our models rely on the mass and metallicity dependencies of planet occurrence rates derived from the literature. In this way, we are able to extrapolate exoplanet population properties to different galactic environments. As expected, the majority of exoplanets are located in planetary systems around M stars (see Fig. \ref{fig:teff_instellation}) because these are the most abundant stars, although they are observable only in very close-by systems.

To further validate and test our modelling, we also create a full forward simulation of the {\it Kepler} satellite's census of exoplanets (Sect. \ref{sec:kepler}, Figs. \ref{fig:kepler_field_hist_magn} -- \ref{fig:kepler_period_radius}, Appendix \ref{sec:annexe_kepler}). Our modelling relies on the same galaxy simulation and includes detailed prescriptions for {\it Kepler's} observing campaign. This experiment resulted in a comparable number of exoplanets compared to the observed {\it Kepler} census, with a similar distribution as \textit{Kepler} for giants, HJ and CJ (see Fig. \ref{fig:kepler_period_radius}). However, the {\it Kepler} modelling also demonstrated several shortcomings of our modelling that will be addressed in the future (see below).

When extending our modelling to different regions of the same galaxy (Fig. \ref{fig:g755_regions_summary}), we find that thick disc/halo, and outer-disc regions host less planets per star than the SN and other more central regions of the galaxy\footnote{Exception made of the \texttt{centre} region, which exhibit a secondary metallicity pick at lower metallicity ($\sim -0.3$).}, as well as more Earth-like planets in proportion. The main driver for this is the metallicity dependence of the planet occurrence rates.
Comparing the SN equivalent in different galaxies of the simulation suite (Fig. \ref{fig:6galaxies_summary}), we again find that the main factor of difference in planet type occurrence rates is the metallicity distribution of the stellar population. More metal-poor galaxies have a smaller number of planets per star, and host fewer giant planets, but (in percentage) more Earth-like.

In total, we obtain $\sim6.7$ planets per star, higher than the \cite{Boettner2024} estimate ($\sim5.6$) or the one of \cite{Hsu2019} ($\sim5.0$). As discussed below, our higher estimate could be influenced by some of the approximations used here.

\subsection{Caveats and future work}\label{sec:caveats}

We recall the several limitations in our analysis that should be addressed in future work:
\begin{itemize}
    \item Our sample excludes planets in binary or higher-order star systems, which may influence the overall occurrence rates.
    \item Exoplanet formation yields are strongly metallicity-dependent, yet the precise relationship between occurrence rates and stellar [Fe/H] remains poorly constrained. Improved models incorporating the joint dependence of planet occurrence on stellar mass and metallicity are needed. In particular, we assumed very conservative limits for the occurrence rate below [Fe/H] $=-1$, which will only be improved by future space missions like PLATO and HAYDN.
    \item The conversion from planetary mass to radius introduces significant uncertainty due to the degeneracy in composition and structure models.
    \item By defining planet types based on mass rather than originally assigned planet types, we inherently modify the original occurrence rates. In the future, we will improve this categorisation by combining masses and radii to define the planetary types by density. This will allow us to obtain a finer division between Earth-like planets, super-Earths, water worlds, sub-Neptunes, Neptunes, etc.
    \item Our model does not account for orbital resonances or system architecture, which may affect the detectability and distribution of planets.
    \item We do not incorporate the observed radius gap nor the Neptune desert, which requires a clearer distinction between super-Earths and sub-Neptunes in future work. We plan to incorporate different photo-evaporation effects in a future version of our exoplanet simulation.
\end{itemize}

Our full forward simulation of the {\it Kepler} exoplanet census revealed some additional caveats:
\begin{itemize}
    \item The underlying galaxy simulation {\tt g7.55e11} shows some important differences with the Milky Way in terms of star-formation history (the simulation has too many young stars) and disc structure (the thin disc is kinematically too hot) that are reflected in systematic differences also in the stellar population of the {\it Kepler} field. In the future we will therefore also apply our method to Milky-Way-like star-count simulations like TRILEGAL \citep{Girardi2005, DalTio2021} or the Besançon Galaxy Model \citep{Robin2003, Robin2022, delAlcazar-Julia2025}.
    One of our goals will be to compare the exoplanet population of the different galactic components (like \citealt{Boettner2024b}).
    \item The model overproduces detectable exoplanets relative to the observed number of stars, potentially suggesting a potential overestimation of the underlying occurrence rates.
    \item The ``detectable'' exoplanets we obtain in the \textit{Kepler} field do not occupy the same regions of the radius-period space as the observed one. In addition to the over-population of the radius gap and Neptune desert, we under-populate the region of sub-Earth radii ($R<1~R_{\oplus}$) as well as super-Earths with periods over 100 days. Those differences could be explained by the important differences in the parameters of the host stellar population, especially stellar radii.
    \item The synthetic stellar sample from the galactic simulation does not replicate the \textit{Kepler} Input Catalogue well enough, limiting the accuracy of our comparison with observed exoplanet demographics. Future work will refine this approach by applying our methodology directly to the \textit{Kepler} Input Catalogue, thus enabling more robust constraints on planet occurrence rates (see, e.g.  \citealt{Matuszewski2023, Boettner2024b}).
\end{itemize}
 
In the future, we plan to improve our exoplanet population simulations by addressing several of the points mentioned above, to be able to apply and test our predictions for the next generation of exoplanet discovery missions, such as PLATO, {\it Roman}, {\it Ariel}, or HAYDN.

\section*{Data availability}

The NIHAO-UHD simulations used in this work were produced by \citet{Buck2020} and are available at \url{https://tobias-buck.de/\#sim_data}. The code used to produce the figures in this paper can be found in at \url{https://github.com/cpadois/glxsimu_to_exoplanets_paper}.

\begin{acknowledgements}
We thank the anonymous referee for their constructive comments which improved greatly the clarity of the paper.
This paper is funded by the Horizon Europe Marie Skłodowska-Curie Actions Doctoral Network MWGaiaDN (Grant agreement No. 101072454; \url{https://www.mwgaiadn.eu/}), co-funded by UK Research and Innovation (EP/X031756/1).
This work was partially funded by the Spanish MICIN/AEI/10.13039/501100011033 and by the ``ERDF A way of making Europe'' funds by the European Union through grant RTI2018-095076-B-C21 and PID2021-122842OB-C21, and the Institute of Cosmos Sciences University of Barcelona (ICCUB, Unidad de Excelencia ’Mar\'{\i}a de Maeztu’) through grant CEX2019-000918-M. FA acknowledges financial support from MCIN/AEI/10.13039/501100011033 through a RYC2021-031638-I grant co-funded by the European Union NextGenerationEU/PRTR. 
JA is supported by the National Natural Science Foundation of China under grant No. 12233001, by the National Key R\&D Program of China under grant No. 2024YFA1611602, by a Shanghai Natural Science Research Grant (24ZR1491200), by the ``111'' project of the Ministry of Education under grant No. B20019, and by the China Manned Space Project with Nos. CMS-CSST-2025-A08, CMS-CSST-2025-A09, and CMS-CSST-2025-A11. 
D.S. acknowledges support from the Foundation for Research and Technological Innovation Support of the State of Sergipe (FAPITEC/SE) and the National Council for Scientific and Technological Development (CNPq), under grant numbers 794017/2013 and 444372/2024-5.
EM acknowledges funding from FAPEMIG under project number APQ-02493-22 and a research productivity grant number 309829/2022-4 awarded by the CNPq.

The preparation of this work has made use of TOPCAT \citep{Taylor2005}, NASA's Astrophysics Data System Bibliographic Services, as well as the open-source Python packages \texttt{astropy} \citep{Astropy2018}, and \texttt{numpy} \citep{VanderWalt2011}. The figures in this paper were produced with \texttt{matplotlib} \citep{Hunter2007}. 

This research has made use of the NASA Exoplanet Archive \url{https://exoplanetarchive.ipac.caltech.edu/}, which is operated by the California Institute of Technology, under contract with the National Aeronautics and Space Administration under the Exoplanet Exploration Program.

This work has made use of data from the European Space Agency (ESA) mission {\it Gaia} (\url{http://www.cosmos.esa.int/gaia}), processed by the {\it Gaia} Data Processing and Analysis Consortium (DPAC, \url{http://www.cosmos.esa.int/web/gaia/dpac/consortium}). Funding for the DPAC has been provided by national institutions, in particular the institutions participating in the {\it Gaia} Multilateral Agreement. \\

\end{acknowledgements}

\bibliographystyle{aa}
\bibliography{biblio_paper_clean}

\begin{appendix}

\section{Method details}\label{sec:annexe_method}

\subsection{Generating single, binary and higher-order stellar systems}
\label{annexe:generate_multiple_stellar_syst}

The multiplicity fraction (MF) is defined as the number of binary or higher-order systems observed for stars of a determined mass. We interpolate the MF and the companion frequency (CF) as a function of the primary mass given by \citealt{Offner2023}. We then iterate over the simulated stars, starting from the highest masses: for each star, we first determine whether it is a single or the primary of a multiple system. If it is non-single, we draw a random number of companions from a Poisson distribution with an expectation value that follows the CF.
The numbers we obtain for the simulated SN are given in Sect. \ref{sec:results_SN}.

For primary stars, we draw a random mass ratio $q$ for their respective number of companions, following a uniform distribution between 0.05 and 1.0. Then we assign a star from the single star population with the closest masses $\mathrm{M}_{\mathrm{compa}} = q ~\mathrm{M}_{\mathrm{prim}}$ as companions. If the drawn $q$ gives a companion mass lower than the sample minimal mass tabulated in the PARSEC models ($0.1 ~\mathrm{M}_{\odot}$), we mark that primary star with a \texttt{low-mass companion} flag. After this process, we have categorised all created stars into three categories: single stars, primary stars in multiple systems, and companion stars.
In this paper, as a first approximation, we discard all stars involved in multiple systems and focus on the single-star population.

\begin{figure}
    \centering
    \includegraphics[width=0.48\textwidth]{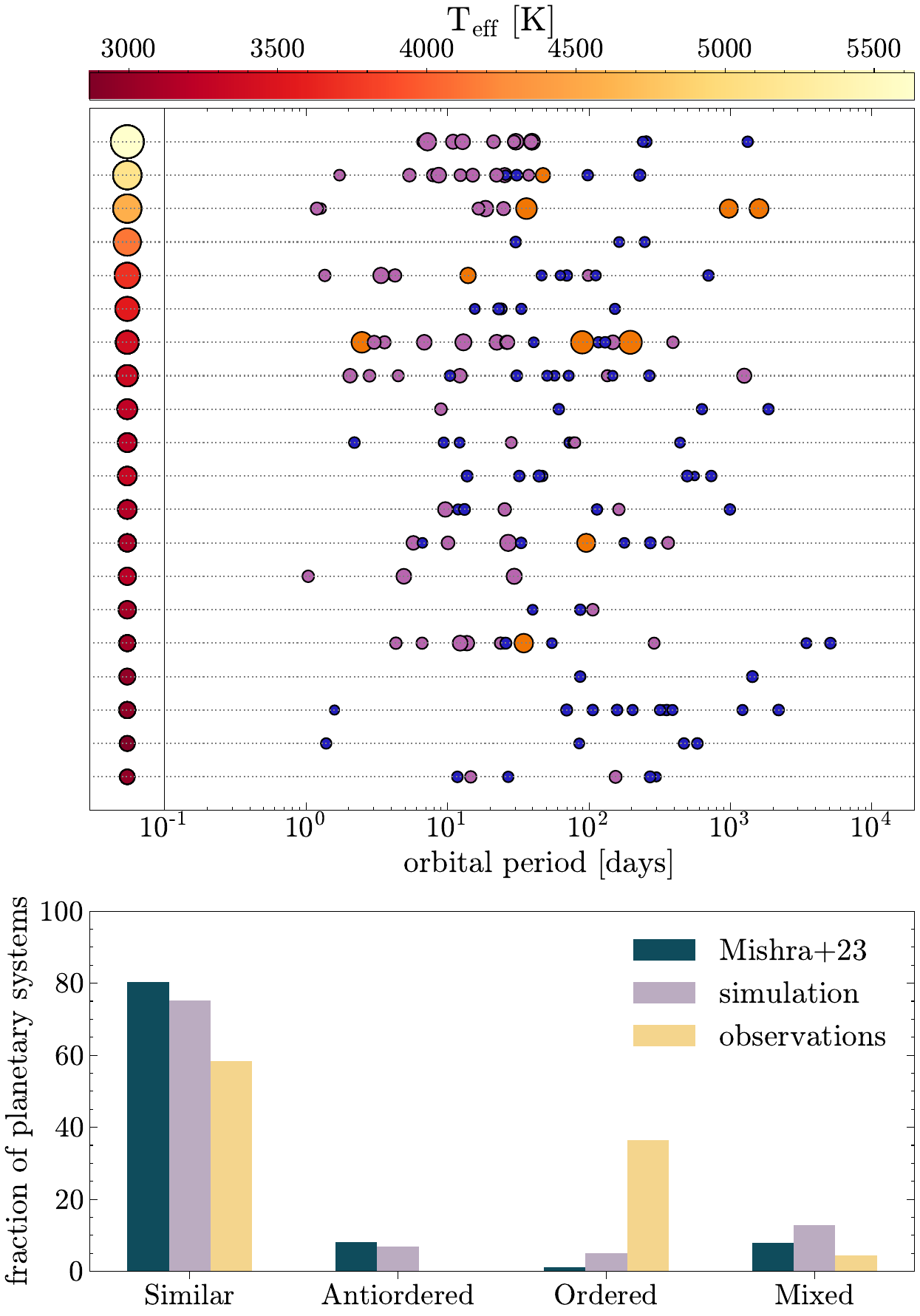}
    \caption{Top panel: Randomly selected synthetic planetary systems, sorted by stellar mass. Stars are coloured by effective temperature, and their size is based on their radius. The planets are coloured as in Fig.\ref{fig:occ_rate_planets_mass_feh}.
    An arbitrary scale is used for stellar radius and planet size for illustrative purposes.
    Bottom panel: Distribution of our generated planetary systems in the different classes defined by \cite{Mishra2021}, compared to their results and to the observations (from their paper too).}
    \label{fig:planetary_systs}
\end{figure}

\subsection{Mass-period distributions and planet types}\label{sec:annexe_mass-period}

In Table \ref{tab:bivariate_gaussian_params} we report the parameters of the bivariate Gaussian distributions used to assign mass and period to our simulated exoplanets (see Sect. \ref{sec:mass-period}). The covariance matrix is defined as $\begin{bmatrix}
  \sigma_{\rm P}^2 & \sigma_{\rm P} ~ \sigma_{\rm M} ~ \rho \\
  \sigma_{\rm P} ~ \sigma_{\rm M} ~ \rho & \sigma_{\rm M}^2
\end{bmatrix}$.

\begin{table}[h!]
\caption{Parameters of the bivariate Gaussian distributions in the mass-period diagram.}
\label{tab:bivariate_gaussian_params}
\centering
\begin{footnotesize}
\begin{tabular}{@{} c c c c c c @{}} 
\hline\hline
    planet type & $\mu_{\rm P}$ [days] & $\mu_{\rm M}$ [$\rm M_{\oplus}$]& $\sigma_{\rm P}$ & $\sigma_{\rm M}$ & $\rho$\\
    \hline
    hot Jupiters (HJ)  & $\log{(4\cdot 10^0)}$ & $\log{(2\cdot 10^2)}$ & $0.2$ & $0.4$ & $-0.6$\\
    cold Jupiters (CJ) & $\log{(5\cdot 10^2)}$ & $\log{(7\cdot 10^2)}$ & $0.6$ & $0.6$ & $0$\\
    SE/Neptunes      & $\log{(1\cdot 10^1})$ & $\log{(1\cdot 10^1)}$ & $0.5$ & $0.3$ & $0.4$\\
    Earth-like         & $\log{(1\cdot 10^2)}$ & $\log{(1\cdot 10^0)}$ & $0.8$ & $0.3$ & $0$\\ \hline
\end{tabular}
\tablefoot{$\mu_{\rm P}$ and $\mu_{\rm M}$ are the means of the period and mass distribution, respectively in days and in Earth mass. $\sigma_{\rm P}$ and $\sigma_{\rm M}$ are the respective dispersions, and $\rho$ is a correlation value.}
\end{footnotesize}
\end{table}

\section{Details of our simulation of the \textit{Kepler} field}\label{sec:annexe_kepler}

\subsection{Comparison of the host stars parameters distribution}
\label{annexe:comparison_stellarparams_hoststars_Kep_fov}

\begin{figure}[h]
    \centering
    \includegraphics[width=0.49\textwidth]{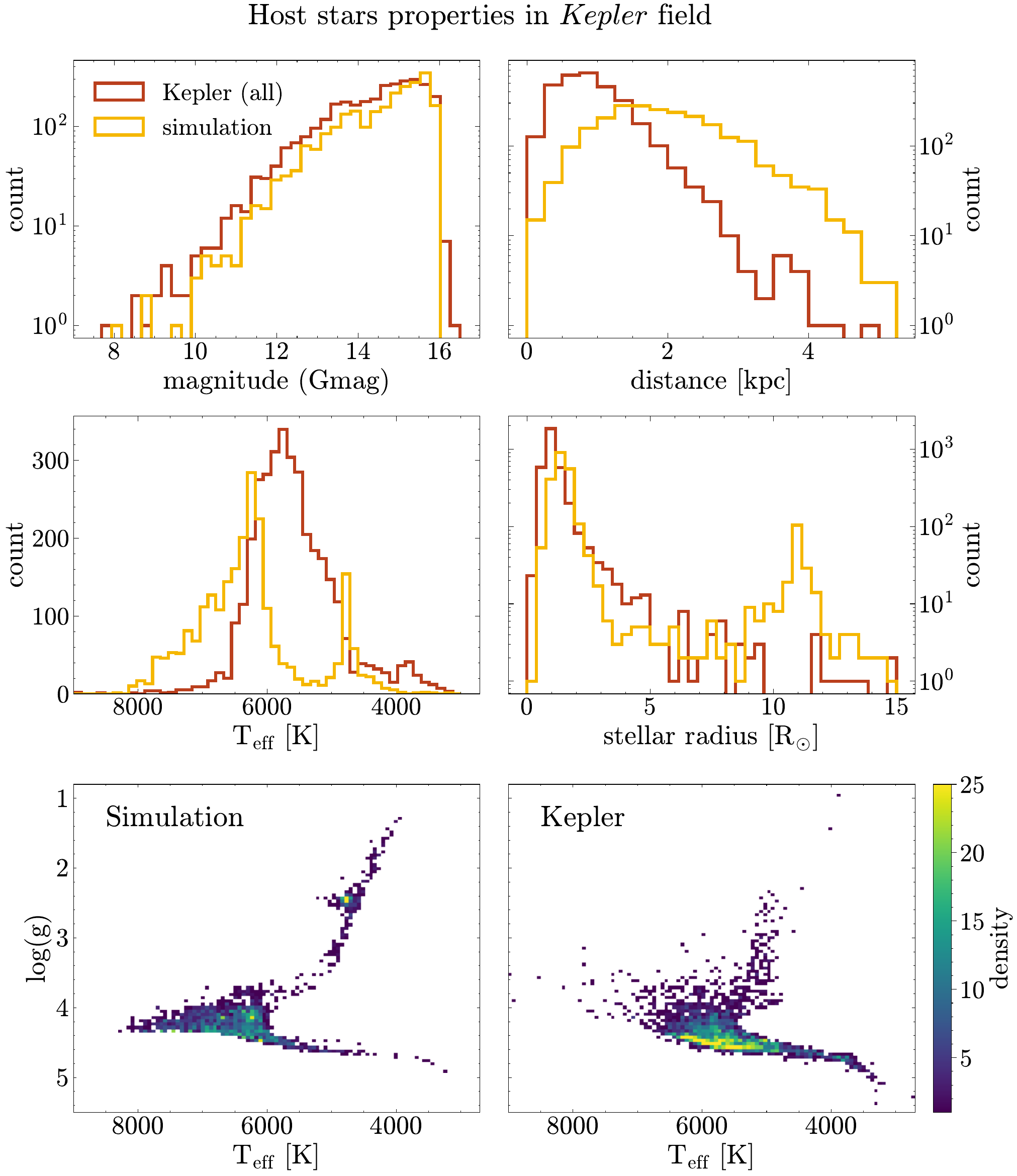}
    \caption{Comparison of host stars properties distribution, between simulated ``detected'' exoplanet host stars (yellow histograms in top panels) and {\it Kepler} confirmed and candidate exoplanet hosts (red). Top and middle row: Magnitude, distance, effective temperature, and stellar radius distributions. Lower panels: {\it Kiel} diagrams of simulated hosts (left) and {\it Kepler} planet hosts (right).}
\label{fig:comparison_stellarparams_hoststars_Kep_fov}
\end{figure}

In Fig. \ref{fig:comparison_stellarparams_hoststars_Kep_fov}, we compare the simulated host star population in the \textit{Kepler} field of view to the detected host stars by \textit{Kepler}. We plot their distribution in magnitude, distance, effective temperature, and stellar radius.
Although the magnitude distributions are coherent (which is expected because we reproduced the {\it Kepler} magnitude distribution in Sect. \ref{sec:selection_fct_kepler}), we see important differences in the distance, effective temperature, and stellar radius distributions. The distribution of those three parameters for the host stars of detectable planets directly reflects the distribution of the total simulated stellar population in the \textit{Kepler} field.

\subsection{Conversion from planets mass to radius}
\label{sec:annexe_mass_to_rad}

The mass-radius relation depends on the type of planet and its composition, but it is not an exact linear relationship. To obtain an accurate global distribution while keeping a simple process of conversion, we used the relations derived by \cite{Parc2024}, adding a Gaussian dispersion to reproduce the observed distribution (see eq. \ref{eq:mass_radius_conversion_parc24}).
We randomly distribute equally planets with masses between $5$ and $14~M_{\oplus}$ between the first two relations.

\begin{multline}
    \label{eq:mass_radius_conversion_parc24}
R= 
\begin{cases}(1.02 \pm 0.08) M^{0.28} & , \text{ if } M<5~M_{\oplus}  \\ 
(0.61 \pm 0.15) M^{0.67} & , \text{ if } 14~M_{\oplus}<M<138~M_{\oplus} \\ 
(11.9 \pm 4) M^{0.01} & , \text{ if } M>138~M_{\oplus}\end{cases}
\end{multline}

\subsection{CDPP estimation}
\label{annexe:CDPP}

To estimate the effective $\rm CDPP$ of each planet, we use the formula of \cite{Christiansen2012}: $\rm CDPP_{eff} = \sqrt{{\rm t_{CDPP}}/{\rm t_{dur}}} \times CDPP_N,$

with $\rm t_{dur}$ being the planet's transit duration in hours, $\rm t_{CDPP}$ the closest provided duration (3h, 6h or 12h, here we use $\rm t_{CDPP}=6$h),
and $\rm CDPP_N$ the root-mean-square (rms) CDPP value associated with the $\rm t_{CDPP}$ at the considered host star's magnitude. We thus scale the tabulated $\rm CDPP_N$ to the actual duration of the transit. To do so, we need to calculate the two quantities $\rm t_{dur}$ and $\rm CDPP_N$, as detailed below.

\vspace{0.3 cm}
\noindent \textbf{Assigning random inclination.}
As a first step, we assign an inclination to every exoplanet's orbit. We assume all planets belonging to a planetary system to have the same orbital inclination. We assign an inclination $i$ to all host stars in our simulated \textit{Kepler} field of view, following a uniform random distribution for $\cos(i)$. This leads to an average percentage of observable transits of around $2.5\%$.

\vspace{0.3 cm}
\noindent \textbf{Calculate $\rm t_{dur}$.}
The transit duration of an exoplanet can be defined in several ways (see \cite{Kipping2010} for an overview). In this work, since we assume all simulated exoplanets to have circular orbits, we chose to use the simplified expression first presented by \cite{Seager2003}, with $\rm P$ the orbital period in hours, $i$ the orbit inclination in degrees and $a_{\rm R}$ the semi-major axis in stellar radius units:
\begin{equation}
    \label{eq:transit_duration}
    {\rm t_{dur}} = \frac{\rm P}{\pi} \arcsin \left( \frac{\sqrt{1 - a_{\rm R}^2 \cos^2(i)}}{a_{\rm R} \sin(i)} \right),
\end{equation}
The computation of $\rm t_{dur}$ is possible only if $\cos(i) \le a_{\rm R}^{-1}$.

\vspace{0.3 cm}
\noindent \textbf{Fitting $\rm CDPP_N$ as a function of host star's magnitude.}
The $\rm CDPP_N$ is usually derived from the lightcurve analysis. We aim to simulate this quantity for our synthetic sample by reproducing the $\rm CDPP_N$ distribution for \textit{Kepler} Quarter 3 planetary targets, given by \cite{Christiansen2012}. We fit the distribution as a function of the magnitude for $\rm t_{dur}=6$h. 
Giant stars ($\log(g) \le 4$) are divided into two groups, one having higher rms CDPP, likely due to higher variability \citep{koch_kepler_2010, Christiansen2012}. We separate dwarfs from giants, sampling giant's CDPP from the upper or lower branch randomly (half in each).

\subsection{Transit depth and S/N comparison for {\it Kepler} confirmed exoplanets} \label{sec:annexe_trdepth_snr_comparison_kepler}

\begin{figure}
    \centering
    \includegraphics[width=0.99\linewidth]{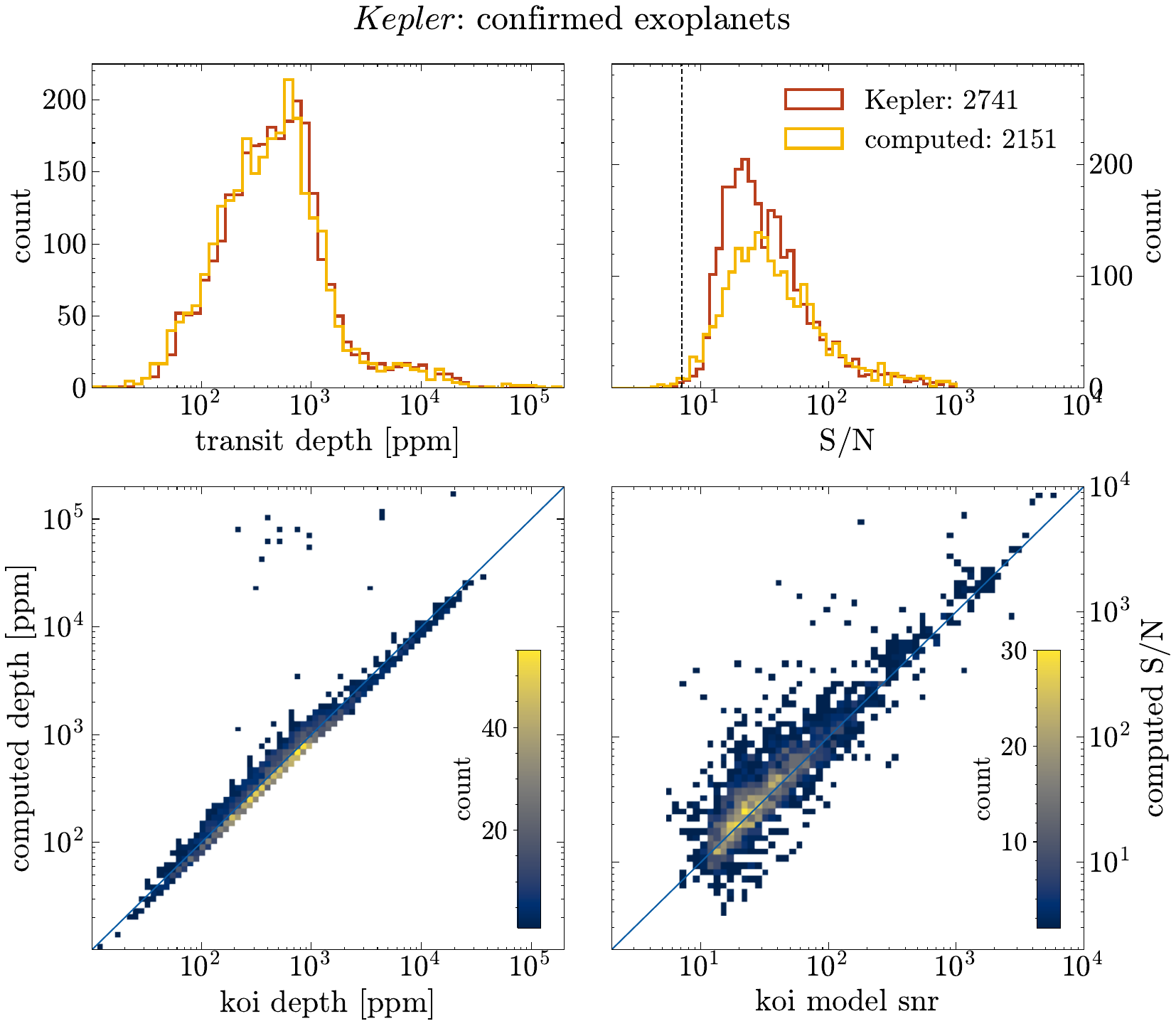}
    \caption{Comparison of calculated vs. observational transit-depth and S/N ratio distributions of {\it Kepler} exoplanets. Top panels: histograms (calculated: yellow, KOI-reported: red). Bottom panels: one-to-one comparisons.}
\label{fig:comparaison_SNR_Kepler}
\end{figure}

To evaluate the accuracy of our S/N estimates, we run the calculations detailed in Sect. \ref{annexe:CDPP} for the planets detected and confirmed by \textit{Kepler}. In Fig.~\ref{fig:comparaison_SNR_Kepler} we compare our results to the published values of \textit{Kepler} Objects of Interest catalogue (KOIs). The transit depth, calculated as the square ratio of the planet's and the host star's radius, matches well the observational \texttt{koi\_depth}. The median logarithmic difference between computed and {\tt koi\_depth} is $-0.067$ dex, with an absolute deviation of $0.024$ dex.
The planet detection S/N is obtained using equation~\ref{eq:SNR_christiansen12}. The median of $\log_{10}$(computed $S/N) - \log_{10}$({\tt koi\_model\_snr}) is $0.026$ dex, with an absolute dispersion of $0.094$ dex. We stress that the S/N is not computed for all \textit{Kepler} exoplanets, because $\sim 20$\% have an inclination not compatible with our transit duration estimation (see equation \ref{eq:transit_duration}).

\end{appendix}

\end{document}